\documentclass[a4paper,11pt]{article}
\pdfoutput=1 

\usepackage{jheppub1} 

\usepackage[latin3]{inputenc}
\usepackage[makeroom]{cancel}
\usepackage{graphicx}
\usepackage{amsmath}
\usepackage{amsfonts}
\usepackage{amssymb}
\usepackage{color}
\usepackage[export]{adjustbox}
\usepackage{graphicx}
\usepackage{dcolumn}
\usepackage{bm}
\usepackage{simplewick}
\usepackage{array}
\usepackage{appendix}
\usepackage{relsize}
\usepackage{subcaption}
\usepackage{float}
\notoc

\newcommand{\be}{\begin{equation}}
	\newcommand{\ee}{\end{equation}}
\newcommand{\beq}{\begin{equation}}
	\newcommand{\eeq}{\end{equation}}
\newcommand{\bea}{\begin{eqnarray}}
	\newcommand{\eea}{\end{eqnarray}}

\title{\boldmath Exotic Lovelock black holes and extended quasitopological electromagnetism}

\author[1]{Askar Ali}

\author[2,3]{and Khalid Saifullah}

\affiliation[1]{Department of Sciences and Humanities, National University of Computer and Emerging Sciences, Peshawar 25000, Pakistan}
\affiliation[2]{Department of Mathematics, Quaid-i-Azam University, Islamabad, Pakistan}
\affiliation[3]{School of Mathematical Sciences, Queen Mary University of London, London, United Kingdom}

\emailAdd{askarali@math.qau.edu.pk}\emailAdd{saifullah@qau.edu.pk}

\abstract{The generalization of Birkhoff's theorem for higher dimensions in Lovelock gravity permits us to investigate the black hole solutions with horizon geometries of nonconstant curvature. We present a new class of exotic dyonic black holes in the context of Lovelock gravity and generalized quasitopological electromagnetism. First, we derive the polynomial equation that describes exotic dyonic black holes in Lovelock gravity with an arbitrary order. Next, the solutions that characterize dyonic exotic black holes of the Gauss-Bonnet and third order Lovelock gravities are worked out. Then we compute the basic thermodynamic quantities for these exotic dyonic black holes. It is also verified that these quantities satisfy the generalized first law and Smarr's relation. Furthermore, the impact of generalized quasitopological electromagnetism and topological parameters on the local stability of the resulting objects are also investigated.
\vspace{80 mm}
}

\notoc 

\begin{document}
\maketitle
\flushbottom 


\section{Background}
\label{sec:intro}
The prevalent theory of gravity in four dimensions is Einstein's theory of gravity (ETG), which explains our Universe at both medium and large scale measurements. A century after Einstein's noteworthy assumptions, the widely recognized discovery of gravitational waves provides credibility to this model. Its adjustment is still required, however, as one would not expect it to hold at exceptionally high energies approaching the Planck scale. Addressing the effects of gravity in higher dimensions seems effective because string theory \cite{1a} and brane cosmology \cite{1b,1c} are believed to provide reliable predictions about the possibility of higher dimensions. Lovelock theory \cite{1} proposes an overarching and extensive framework to explore gravity in this situation. The action of this theory incorporates terms exhibiting higher curvature, and in four dimensions, the Einstein-Hilbert action can be extracted from it. The field equations that follow from the variation of the Lovelock action have been shown to be of second order in the metric derivatives. This demonstrates a ghost-free quantization of the linearized Lovelock gravity \cite{2}. Lovelock gravity has attracted a great deal of theoretical attention, as these higher curvature corrections to ETG unavoidably emerge in quantum gravity. The Gauss-Bonnet term, for example, serves as a leading correction in the ten dimensional gauged super-gravity, while it becomes apparent as a quantum correction for heterotic string theory \cite{2,4}. Higher curvature theories, such as Lovelock theories, share an underlying characteristic that the black hole's entropy no longer remains proportional to the horizon area. Hence, these theories offer an interesting aspect of investigation into black hole thermodynamics as they propose novel insight into how the effects of quantum gravity might alter the radiative behaviour of black holes \cite{4a,4b}. The definition of the Lovelock action is as follows:
\begin{equation}
\mathfrak{I}_g=\int d^dx\sqrt{-g}\sum_{p=0}^{p_{max}}a_p\mathfrak{L}_p, \label{1s}
\end{equation}   
where $\mathfrak{L}_p$'s illustrate the Euler densities for the $2p$-dimensional manifolds so that
\begin{equation}
\mathfrak{L}_p=\frac{1}{2^p}\delta^{\mu_1\nu_1...\mu_{p}\nu_{p}}_{\sigma_1\lambda_1...\sigma_{p}\lambda_{p}}R^{\sigma_1\lambda_1}_{\mu_1\nu_1}...R^{\sigma_p\lambda_p}_{\mu_p\nu_p}.\label{2s}
\end{equation} 
Keep in consideration that $p_{max}=[\frac{d-1}{2}]$ in which the brackets indicate the integer portion of $(d-1)/2$, and $a_p$'s signify the arbitrary coupling parameters. Additionally, $\delta^{\alpha_1\beta_1...\alpha_{p}\beta_{p}}_{\kappa_1\rho_1...\kappa_{p}\rho_{p}}$ is the generalized Kronecker delta, $g_{\sigma\beta}$ denotes the metric tensor and $R^{\rho\lambda}_{\sigma\beta}$ designate the Riemann tensor components. Remember that the nontrivial field equations require the condition $d>2p_{max}$.  
 
 Black holes are remarkable structures that can be encountered in any model of gravity and they play a crucial role in testing these models. These objects additionally feature thermodynamic characteristics that could provide a glimpse into some aspects of the description of quantum gravity. The significance of these objects from different aspects is evident from the fact that various Lovelock black holes and their thermodynamic properties have been thoroughly examined in literature (see, for example, \cite{4b,4c,4d,5,6,6a,7,7f1,7f2,7f3,7f4,7f5,7f6,7f7,7f8,7f9,7f10,7f11,7f11a,7f12,7f13,7f14,7f15,7f16,7f17,7f18}).
 
  Until recently, investigations of the physical characteristics of Lovelock black holes have only looked at the black hole solutions with a base manifold of constant curvature. However, this restriction is not necessary and the investigation of black hole solutions that possess nonconstant curvature horizons could provide very interesting insights. For example, Page presented the first explicit inhomogeneous four-dimensional compact Einstein metric \cite{8f1}. Similarly, the extension of this solution in higher dimensions was also obtained \cite{8f2}. In addition, Bohm also found a class of inhomogeneous solutions with positive curvature on product of spheres \cite{8f3}. The physical properties of these inhomogeneous metrics were studied in Ref. \cite{8f4}. Recently, a more general class of black hole solutions whose transverse space is a nonconstant curvature manifold has also been discovered in the framework of Lovelock gravity \cite{8f5}. These solutions take the form of a warped product of a transverse base manifold and a two-dimensional space \cite{8f6,8f7,8f8,8f9,8f10,8f11,8f12,8f13,8f14}. In four dimensions, Birkhoff's theorem implies that the Schwarzschild metric represents a unique spherically symmetric vacuum solution with positive curvature, whereas Schwarzschild-Tangherlini is the corresponding solution in higher dimensions. When the cosmological constant $\Lambda\neq 0$, the Kottler metric plays the same role. In case an ansatz for field equations is chosen for an arbitrary base manifold, the equations of Einstein's theory require that this base manifold would be the Einstein manifold. In Lovelock gravity, the generalization of Birkhoff's theorem implies that the solutions having plane, spherical, or hyperbolic symmetry are locally isometric to the associated static solutions of Lovelock gravity \cite{8f15,8f16}. In addition, the field equations constrain the nontrivial intrinsic Lovelock tensors of the base manifold to be arbitrary constants. On the basis of this restriction, the Lovelock black hole solutions with nonconstant curvature base manifolds have been found in the literature. The objects described by these solutions are called exotic Lovelock black holes. Recently, the exotic black holes of Gauss-Bonnet and third order Lovelock gravity theories were also investigated in presence of the Maxwell electrodynamics \cite{9f1,9f2,9f3}. Additionally, the effects of nonlinear electromagnetic field on these Lovelock black holes have also been investigated \cite{9f4,9f5}. In this paper we will investigate the exotic dyonic black holes of Lovelock gravity.         

Lovelock gravity has been formulated from the generalization of a $2p$-dimensional topological invariants in higher dimensions. Hence, we encounter higher curvature terms in the Lagrangian (\ref{2s}) of this theory. In the backdrop of Abelian gauge theories, similar concepts have also been explored, for example, Liu et al. introduced the model of quasitopological (QT) electromagnetism \cite{9f6}. In this framework, the familiar Maxwellian model is modified by incorporating new terms into the Lagrangian of Maxwell's theory. These additional terms are developed from electromagnetic $2$-form $F_{[2]}=dA_{[1]}$ and the metric tensor. Additionally, they have a particular connection with the topological invariants. Some of these terms are polynomials such as
\begin{equation}
V_{[2p]}=F_{[2]}\wedge F_{[2]}\wedge...\wedge F_{[2]}. \label{5a}
\end{equation}        
Note that Eq. (\ref{5a}) is analogous to the Pontryagin densities, and their integrals are completely topological in even dimensions. Conversely, in arbitrary dimensions, these $2p$-forms can be configured to influence the system's dynamics. One way to accomplish this is to consider the squared norm as 
\begin{equation}
U^{(p)}_{[d]}\sim\left|V_{[2p]}\right|^2\sim V_{[2p]}\wedge \ast V_{[2p]}. \label{6a} 
\end{equation}

The normal kinetic term of Maxwell's theory refers to the situation of $p=1$ in Eq. (\ref{6a}). Usually, these invariants contribute non-vanishing terms to the equations of motion. Note that this theory has been referred to as ``QT electromagnetism'' since it has two features. First, its constituents have topological origin, i.e., the forms $V_{[2p]}$. Second, the spectrum of the static solutions originating from pure electric or pure magnetic sources aligns with the respective spectrum of the associated Maxwellian solutions. However, amazing phenomena emerge when one looks into dyons. These models have recently been improved in an approach that incorporates the Abelian gauge field $A_{[1]}$ along with the higher-rank $(p-1)$-form field $B_{[p-1]}$ \cite{9f7}. The field strength responsible for $B_{[p-1]}$ is characterized via $H_{[p]}=dB_{[p-1]}$. Depending on the specific context, this emerging field may have numerous physical perceptions. For example, it might imitate the higher-rank fields that are showing up in string theory, including the $3$-forms of $11$- dimensional super-gravity, the Ramond-Ramond $p$-forms, or the elementary Kalb-Ramond $2$-form $B_{[2]}$. Moreover, it turns out that the Maxwell field when coupled to the $p$-forms facilitates the setting up of a greater variety of black hole systems compared to those identified in Ref. \cite{9f6}. Adopting the extended model of QT electromagnetism \cite{9f7,9f8}, the dyonic hairy black holes of Lovelock-scalar gravity were addressed in Ref. \cite{9f9}. Moreover, investigations have been conducted on the thermodynamic properties, thermal fluctuations, and shadow cast of the black holes of massive gravity within the wider context of this extended model \cite{9f10}. Thus, the ultimate objective of this paper is to introduce the new exotic dyonic black hole solutions of Lovelock gravity with extended QT electromagnetism.

  The sequence of our paper is as follows. We begin the second section by reviewing the extended model of QT electromagnetism and Lovelock gravity. Then we establish the polynomial equation that allows for the novel exotic dyonic black hole solutions of Lovelock gravity. In Section 3, we focus on the exotic black holes of Gauss-Bonnet gravity sourced by extended QT electromagnetism. Likewise, the exotic dyonic black holes of the third order Lovelock gravity are addressed in Section 4. Finally, we have a summary and conclusion of our entire paper in Section 5.

\section{Exotic dyonic black holes of Lovelock gravity} 

The action characterizing the Lovelock gravity sourced by the extended QT electromagnetic field in $d$-dimensions can be presented as
\begin{equation} \begin{split}
\mathfrak{I}&=\int d^dx\sqrt{-g}\bigg[\sum_{p=0}^{p_{max}}\frac{a_p}{2^p}\delta^{\mu_1\nu_1\cdots\mu_{p}\nu_{p}}_{\sigma_1\lambda_1\cdots\sigma_{p}\lambda_p} R^{\sigma_1\lambda_1}_{\mu_1\nu_1}\cdots R^{\sigma_{p}\lambda_{p}}_{\mu_{p}\nu_{p}}+\mathfrak{L}_{qt}\bigg],
\label{7a}\end{split}
\end{equation}
whereby the Lagrangian density of extended QT electromagnetism is identified by $\mathfrak{L}_{qt}$. In addition, $a_p$'s are arbitrary Lovelock parameters.
One can utilize the action principle to construct the gravitational field equations from (\ref{7a}) as
\begin{equation}
\sum_{p=0}^{p_{max}}\frac{a_p}{2^{p+1}}\delta^{\alpha\sigma_1\lambda_1\cdots\sigma_p\lambda_{p}}_{\beta\kappa_1\rho_1\cdots\kappa_p\rho_{p}} R^{\kappa_1\rho_1}_{\sigma_1\lambda_1}\cdots R^{\kappa_{p}\rho_{p}}_{\sigma_{p}\lambda_{p}}=-\mathfrak{T}^{(M)\alpha}_{\beta}.
\label{9a}
\end{equation}
At this point the energy-momentum tensor $\mathfrak{T}^{(qt)\alpha}_{\beta}$ of the matter source of gravity is offered by
\begin{equation}
\mathfrak{T}^{(M)}_{\alpha\beta}=-\frac{2}{\sqrt{-g}}\frac{\delta \mathfrak{I}_{qt}}{\delta g^{\alpha\beta}},
\label{10a}
\end{equation}
where the $\mathfrak{I}_{qt}$ specifies the action of matter source.
The theoretical framework of extended QT electromagnetism was previously adopted in Refs. \cite{9f7,9f8,9f9,9f10} and structures inspired by (\ref{6a}) are developed with the field strength $H_{[q]}$. In specific, it is reasonable to presume
\begin{align}\begin{split}
&\digamma_{[2p]}=F_{[2]}\wedge F_{[2]}\wedge\cdots\wedge F_{[2]}, p\leq \lfloor d/2\rfloor,\\&
\mathcal{H}_{[sp]}=H_{[s]}\wedge H_{[s]}\wedge\cdots\wedge H_{[s]}, p\leq\lfloor d/s\rfloor,\\&\digamma\mathcal{H}_{[2p+sm]}=\digamma_{[2p]}\wedge\mathcal{H}_{sm}, 2p+sm\leq d.\label{13a}\end{split}
\end{align}
Continuing from these, squared norms are further expressed by means of the Hodge product; these incorporate $\left|\digamma_{[2p]}\right|^2$, $\left|\mathcal{ H}_{[sp]}\right|^2$, and $\left|\digamma\mathcal{ H}_{[2p+sm]}\right|^2$, respectively \cite{9f7}. Undoubtedly, these squared norms are not the only alternatives available. Thus, one can put forward the expressions such as $\digamma_{[2p]}\wedge\ast\mathcal{ H}_{[sm]}$ with $2p=sm$ and $ p\leq \lfloor d/2\rfloor$, $\digamma_{[2p]}\wedge\ast\digamma\mathcal{ H}_{[2r+sm]}$ with $sm=2(p-r)$ and $ p\leq \lfloor d/2\rfloor$. Similar to this, the invariant $\mathcal{H}_{[sp]}\wedge\ast\digamma\mathcal{ H}_{[2r+sm]}$ with $2r=s(p-m)$ and $ p\leq \lfloor d/s\rfloor$ could be additionally utilized. Since each of these constituents are crucial to the equations of motion, the action function associated with $\mathfrak{L}_{qt}$ must take them into account. 

Nonetheless, we perceive the set-ups as 
  \begin{align}\begin{split}
  &F_{\kappa\eta}\sim h'(r)\delta^{x^0x^1}_{\kappa\eta},\\&
  H_{\beta_1\beta_2\cdots\beta_s}\sim \delta^{x^2\cdots x^d}_{\beta_1\beta_2\cdots\beta_s},\label{14a}\end{split}
  \end{align} 
  where $s=d-2$. Furthermore, $F_{\kappa\eta}$ is solely electric while $H_{\beta_1\beta_2\cdots\beta_s}$ is solely magnetic. Remember that the nonzero terms for the above set-ups are the kinetic terms $\left|\digamma_{[2]}\right|^2\sim F_{\kappa\eta}F^{\kappa\eta}$ and $\left|\mathcal{H}\right|^2$, as well as the interacting term $\left|\digamma\mathcal{ H}_{[d]}\right|^2$. For this purpose, it is vital to lay out the Lagrangian density for the matter source as 
  \begin{equation}
  \mathfrak{L}_{qt}=-\frac{1}{4} F_{\rho\sigma}F^{\rho\sigma}-\frac{1}{2m!}H_{\nu_1\nu_2\cdots\nu_m}H^{\nu_1\nu_2\cdots\nu_m}-\zeta\mathfrak{L}_{int}.\label{15a}
  \end{equation}
  Here $\zeta$ is the coupling parameter, while $\mathfrak{L}_{int}$ signifies the interaction term and is specified as
  \begin{equation}
  \mathfrak{L}_{int}=\delta^{\kappa_1\cdots\kappa_d}_{\eta_1\cdots\eta_d}F_{\kappa_1\kappa_2}H_{\kappa_3\cdots\kappa_d}F^{\eta_1\eta_2}H^{\eta_3\cdots\eta_d}.\label{16a}
  \end{equation}
  Using the action principle, one can acquire the QT electromagnetic field equations by putting Eq. (\ref{15a}) into Eq. (\ref{7a}). Hence, we are left with
  \begin{equation}
  \nabla_{\sigma}F^{\sigma\rho}-4\zeta\delta^{\rho\sigma\mu_1\cdots\mu_s}_{\eta_1\eta_2\cdots\eta_d}H_{\mu_1\cdots\mu_s}\nabla_{\sigma}\big(F^{\eta_1\eta_2}H^{\eta_3\cdots\eta_d}\big)=0,\label{17a}
  \end{equation} 
and
\begin{equation}
\nabla_{\rho}H^{\rho\mu_1\cdots\mu_{s-1}}+2\zeta s!\delta^{\rho\sigma\kappa\mu_1\cdots\mu_{s-1}}_{\eta_1\cdots\eta_d}F_{\rho\sigma}\nabla_{\kappa}\big(F^{\eta_1\eta_2}H^{\eta_3\cdots\eta_d}\big)=0.\label{18a}
\end{equation}
Beyond that, the energy-momentum tensor associated with Eq. (\ref{15a}) can also be demonstrated as follows:
\begin{equation}\begin{split}
\mathfrak{T}^{(M)}_{\kappa\sigma}&=F_{\kappa\rho}F^{\rho}_{\sigma}-\frac{1}{4}g_{\kappa\sigma}F_{\rho\eta}F^{\rho\eta}+\frac{1}{2(s-1)!}H_{\kappa\mu_1\cdots\mu_{s-1}}H^{\mu_1\cdots\mu_{s-1}}_{\sigma}\\&-\frac{1}{2(s!)^2}\delta^{\mu_1\cdots\mu_s\rho}_{\eta_1\cdots\eta_s(\kappa}g_{\sigma)\rho}H_{\mu_1\cdots\mu_s}H^{\eta_1\cdots\eta_s}+\zeta g_{\kappa\sigma}\mathfrak{L}_{int}.
\label{19a}\end{split}
\end{equation}

Here the exotic dyonic black hole solutions of Eq. (\ref{9a}) are the focus of our interest. To go about this, let us select a subsequent metric ansatz
 \begin{equation}
 ds^2=-f(r)dt^2+\frac{dr^2}{f(r)}+r^2d\Omega^{2}_{d_2}.
 \label{20a}
 \end{equation}
 For convenience, take note that we are employing the symbol $d_j=d-j$. The functioning of the base manifold, which is illustrated by $d\Omega^{2}_{d_2}$, also explains the importance of this specific metric. For example, offering $d\Omega^{2}_{d_2}$ as the metric of the $d_2$-dimensional hyper-surface with constant curvature $d_2d_3\xi$ is the most frequent case. In this situation, it takes the form
 \begin{equation}\begin{split}
 d\Omega^2_{d_2}=\left\{ \begin{array}{rcl}
 d\theta_1^2+\sum_{j=2}^{d-2}\prod_{l=1}^{j-1}\sin^2\theta_l d\theta_j^2, & 
 & \xi=1, \\d\theta_1^2+\sinh^2\theta_1d\theta_2^2+\sinh^2\theta_1\sum_{j=3}^{d-2}\prod_{l=2}^{j-1}\sin^2\theta_ld\theta^2_j, &  & \xi=-1,\\\sum_{j=1}^{d-2}d\phi_j^2, & & \xi=0. 
 \end{array}\right.\label{21a}
 \end{split}
 \end{equation}
 Notice that the curvatures of hyperbolic, flat, and spherical hyper-surfaces line up with the values $\xi=-1, 0, +1$, correspondingly. However, one might believe that $d\Omega^{2}_{d_2}$ represent a line element of an extended $d_2$-dimensional base manifold that features a nonconstant curvature. Hence, by utilizing the metric (\ref{20a}) in Eqs. (\ref{9a}), a pair of simplified equations for the $d$-dimensional spacetime with a nonconstant curvature base manifold are obtained as 
 \begin{eqnarray}\begin{split}
 	\frac{d_2d_1\delta^{i}_{j}}{2d_1!r^{d_2}}\sum_{p=0}^{p_{max}}\big[(d_{2n+2})!\tilde{\mathfrak{L}}^p\big]\bigg[\frac{d}{dr}\bigg(r^{d_{2n+1}}W_p\bigg(\frac{-f(r)}{r^2}\bigg)\bigg)\bigg]=-8\pi G_N \mathfrak{T}^{(M)i}_{j}	,\label{22a1}\end{split}
 \end{eqnarray}
and
 \begin{eqnarray}\begin{split}
		\frac{d_2d_1}{d_1!r^{d_3}}\sum_{p=0}^{p_{max}}\big[(d_{2n+3})!\tilde{\mathcal{G}}^{(p)\rho}_{\sigma}\big]\bigg[\frac{d^2}{dr^2}\bigg(r^{d_{2n+1}}W_p\bigg(\frac{-f(r)}{r^2}\bigg)\bigg)\bigg]=8\pi G_N \mathfrak{T}^{(M)\rho}_{\sigma}.\label{23a1}\end{split}
\end{eqnarray}
 Remember that both of $i,j$ take the values $0$ and $1$, whereas $W_p$ is the polynomial specified by
 \begin{equation}
 W_p\bigg(\frac{-f(r)}{r^2}\bigg)=\sum_{k=p}^{p_{max}}\alpha_{k}\binom{k}{p}\bigg(\frac{-f(r)}{r^2}\bigg)^{k-p},\label{24a1}
 \end{equation} 
with $\alpha_{k}$'s are called re-scaled Lovelock coupling parameters, so
\begin{align}\begin{split}
		&\alpha_0=\frac{a_0}{d_1d_2}=-\frac{2\Lambda}{d_1d_2},\\&
		\alpha_1=a_1,\\& \alpha_p=a_p\prod_{n=3}^{2p}(d-n)   \text{      for $k\geq 2$} .\label{24a2}\end{split}
\end{align}
 Moreover, the polynomial (\ref{24a1}) fulfills the recurrence relation
 \begin{equation}
 	\frac{d}{dr}W_p\bigg(\frac{-f(r)}{r^2}\bigg)=(p+1)W_{p+1}\bigg(\frac{-f(r)}{r^2}\bigg).\label{25a1}
 \end{equation} 
  The Euler characteristic and Lovelock tensor of the base manifold are signified by $\tilde{\mathfrak{L}}^p$ and $\tilde{\mathcal{G}}^{(p)\rho}_{\sigma}$, respectively and are specified by
  \begin{align}\begin{split}
  		&\tilde{\mathfrak{L}}^p=\frac{d_2!\beta_p}{d_{2p+2}!},\\&
  	\tilde{\mathcal{G}}^{(p)\rho}_{\sigma}=-\frac{d_3!\beta_p}{2d_{2p+3}!}\delta^{\rho}_{\sigma}.\label{26a1}\end{split}
  \end{align}
 These quantities simplify Eqs. (\ref{22a1}) and (\ref{23a1}) into
  \begin{eqnarray}\begin{split}
  		\frac{d_2\delta^{i}_{j}}{2r^{d_2}}\frac{d}{dr}\sum_{p=0}^{p_{max}}\bigg[\beta_p\bigg(r^{d_{2p+1}}W_p\bigg(\frac{-f(r)}{r^2}\bigg)\bigg)\bigg]=-8\pi G_N \mathfrak{T}^{(M)i}_{j},\label{27a1}\end{split}
  \end{eqnarray}
  and
   \begin{eqnarray}\begin{split}
  		\frac{\delta^{\rho}_{\sigma}}{2r^{d_3}}\frac{d^2}{dr^2}\sum_{p=0}^{p_{max}}\bigg[\beta_p\bigg(r^{d_{2p+1}}W_p\bigg(\frac{-f(r)}{r^2}\bigg)\bigg)\bigg]=-8\pi G_N \mathfrak{T}^{(M)\rho}_{\sigma}.\label{28a1}\end{split}
  \end{eqnarray}
    If the $d_2$-dimensional hyper-surface is embraced with a magnetic field according to its intrinsic volume form, then one may consider
 \begin{equation}
 H_{\nu_1\nu_2\cdots\nu_m}=q\sqrt{\Sigma_{\beta}}\delta^{x^1\cdots x^m}_{\nu_1\cdots\nu_m},\label{29a1}
 \end{equation} 
where $\Sigma_{\beta}$ reflects the volume of the $d_2$-dimensional hyper-surface. Likewise, in the entirely electric case, the Maxwell tensor acquires an analogous form 
 \begin{equation}
 F_{\alpha\beta}=\frac{dh}{dr}\delta^{tr}_{\alpha\beta}. \label{30a1}
 \end{equation}
 Thus, by utilizing Eqs. (\ref{29a1}) and (\ref{30a1}), one gets 
 \begin{equation}
 r^{2d_2}\bigg(d_2\frac{dh}{dr}+r\frac{d^2h}{dr^2}\bigg)-8\zeta(d_2!)^2q^2\bigg(d_2\frac{dh}{dr}-r\frac{d^2h}{dr^2}\bigg)=0.\label{31a1}
 \end{equation} 
 When Eq. (\ref{31a1}) is integrated, it yields
 \begin{equation}
 \frac{dh}{dr}=\frac{e r^{d_2}}{r^{2d_2}+8\zeta(d_2!)^2q^2}.\label{32a1}
 \end{equation}
 It is important to remember that the integration constants $q$ and $e$ correspond to magnetic and electric charges, respectively. Eq. (\ref{32a1}) further specifies the screening of the electric field that is generated due to the interaction of electric field with the analogous magnetic component. Additionally, if we utilize Eqs. (\ref{19a}) and (\ref{32a1}) in the simplified field equations (\ref{27a1}) and (\ref{28a1}), the polynomial equation in $f(r)$ can be acquired as follows: 
 \begin{eqnarray}\begin{split}
 \sum_{p=0}^{p_{max}}\frac{\beta_{p}}{r^{2p}}\bigg[\sum_{k=p}^{p_{max}}\alpha_k\binom{k}{p}\bigg(-\frac{f(r)}{r^2}\bigg)^{k-p}\bigg]&=\frac{16\pi G_NM}{d_2\Sigma_{\beta}r^{d_1}}-\frac{8\pi G_Nq^2}{d_2d_3r^{2d_2}}-\frac{8\pi G_Ne^2}{d_2r^{2d_2}}\\&\times F_1\bigg(\bigg[1,\frac{d_3}{2d_2}\bigg],\bigg[\frac{3d_2-1}{2d_2}\bigg],\frac{-8\zeta q^2(\Gamma(d_1))^2}{r^{2d_2}}\bigg),\label{33a1}\end{split}
 \end{eqnarray}
 where $F_1$ symbolizes the hypergeometric function and the constant of integration $\mu$ is linked to the mass of the black hole $M$ via formula
 \begin{eqnarray}
 M=\frac{(d-2)\mu\Sigma_{\beta}}{16\pi G_N}.\label{34a1}
 \end{eqnarray}
 The solutions which originate from the roots of the polynomial equation (\ref{33a1}) in a way that they vanish at least once for some positive value of coordinate $r$ will reflect the gravitating objects called exotic dyonic black holes of Lovelock gravity sourced by extended QT electromagnetism. Here, the constants $\beta_p$'s have been referred to as topological parameters. One can take $\beta_0=1$ without any sacrifice of clarity. When the $d_2$-dimensional base manifold is of constant curvature $d_2d_3\xi$, i.e. $d\Omega^{2}_{d_2}$ is expressed by Eq. (\ref{21a}), then the polynomial equation (\ref{33a1}) simplifies to the form
  \begin{eqnarray}\begin{split}
 		\sum_{p=0}^{p_{max}}\bigg[\alpha_p\bigg(\frac{\xi-f(r)}{r^2}\bigg)^{p}\bigg]&=\frac{16\pi G_NM}{d_2\Sigma_{\beta}r^{d_1}}-\frac{8\pi G_Nq^2}{d_2d_3r^{2d_2}}-\frac{8\pi G_Ne^2}{d_2r^{2d_2}}\\&\times F_1\bigg(\bigg[1,\frac{d_3}{2d_2}\bigg],\bigg[\frac{3d_2-1}{2d_2}\bigg],\frac{-8\zeta q^2(\Gamma(d_1))^2}{r^{2d_2}}\bigg).\label{35a1}\end{split}
 \end{eqnarray}
   The foregoing polynomial equation (\ref{35a1}) generates the dyonic black holes with maximally symmetric horizons in the scenario of Lovelock gravity and extended QT electromagnetism \cite{9f9}. Notice that the comparison of polynomial equations (\ref{33a1}) and (\ref{35a1}) reveals $\beta_p=\xi^p$ when the base manifolds for dyonic black holes have constant curvature. 
 With regards to Eq. (\ref{20a}), we can reveal the Kretschman scalar as 
 \begin{eqnarray}\begin{split}
 R^{\mu\nu\rho\sigma}R_{\mu\nu\rho\sigma}&=\bigg(\frac{d^2f}{dr^2}\bigg)^2+\frac{2d_2}{r^2}\bigg(\frac{df}{dr}\bigg)^2+\frac{2d_2d_3\big(f(r)\big)^2}{r^4}-4\frac{\mathcal{R}[\Omega]f(r)}{r^4}+\frac{\mathcal{K}[\Omega]}{r^4},\label{36a1}\end{split}
 \end{eqnarray}
where the Ricci and Kretschmann scalars of the base manifold are written as $\mathcal{R}$ and $\mathcal{K}$, respectively. The essential singularity is demonstrated by the non-analytic response of Kretschmann scalar at $r=0$. Hence, we will employ this expression (\ref{36a1}) for investigating the structure of exotic dyonic black hole solutions of Gauss-Bonnet and third order Lovelock gravities in the upcoming sections.   

\section{Exotic dyonic black holes of Gauss-Bonnet gravity}
In this section we investigate the exotic black holes of Gauss-Bonnet theory and their physical properties. The resultant polynomial (\ref{33a1}) for $p_{max}=2$ reveals the solution 
\begin{eqnarray}\begin{split}
f_{\pm}(r)&=\frac{r^2+2\alpha_2\beta_1}{2\alpha_2}\pm\frac{\sqrt{4\alpha_2^2(\beta_1^2-\beta)+r^4(1-4\alpha_2\alpha_0)+F_{GB}(r)}}{2\alpha_2}, \label{37a}\end{split}
\end{eqnarray}
with
\begin{equation}\begin{split}
F_{GB}(r)&=\frac{64\pi G_N\alpha_2M}{d_2\Sigma_{\beta}r^{d_5}}-\frac{32\pi G_N\alpha_2q^2}{d_2d_3r^{2d_4}}-\frac{32\pi G_N\alpha_2e^2}{d_2r^{2d_4}}\\&\times F_1\bigg(\bigg[1,\frac{d_3}{2d_2}\bigg],\bigg[\frac{3d_2-1}{2d_2}\bigg],\frac{-8\zeta q^2(\Gamma(d_1))^2}{r^{2d_2}}\bigg).\label{38a}\end{split}
\end{equation}
 Note that $\alpha_0$ and the cosmological constant $\Lambda$ are linked together through Eq. (\ref{24a2}). When the base manifold has constant curvature and $q=e=\Lambda=0$, then the factually genuine solution within $f_+(r)$ and $f_-(r)$ is the one which matches the Schwarzschild-Tangherlini solution. Consequently, the solution $f_+(r)$ is not physically valid. Nevertheless, the resultant solution $f_-(r)$ exhibits the following asymptotic value in the vicinity of $\alpha_2=0$ 
 \begin{eqnarray}\begin{split}
 		f_{-}(r)&=\beta_1-\frac{2\Lambda r^2}{d_1d_2}-\frac{16\pi G_NM}{d_2\Sigma_{\beta}r^{d_3}}+\frac{8\pi G_Nq^2}{d_2d_3r^{2d_3}}+\frac{8\pi G_Ne^2}{d_2r^{2d_3}}\\&\times F_1\bigg(\bigg[1,\frac{d_3}{2d_2}\bigg],\bigg[\frac{3d_2-1}{2d_2}\bigg],\frac{-8\zeta q^2(\Gamma(d_1))^2}{r^{2d_2}}\bigg). \label{39a}\end{split}
 \end{eqnarray}
This equation is analogous to the dyonic black hole solution of ETG sourced by QT electromagnetism whenever the base manifold has constant curvature, i.e. $\beta_1=\xi$. Additionally, when the electromagnetic charges vanish, the preceding expression leads to the Schwarzschild-Tangherlini solution. However, when one sets $p_{max}=2$ and $8\pi G_N=1$ in Eq. (\ref{35a1}), the solution expressing the dyonic black holes with constant curvature horizons \cite{9f9} can be established as follows:
\begin{eqnarray}\begin{split}
		f_{-}(r)&=\xi+\frac{r^2}{2\alpha_2}\bigg[1- \sqrt{\mathcal{W}(r)}\bigg], \label{40a}\end{split}
\end{eqnarray}
where
\begin{equation}\begin{split}
		\mathcal{W}(r)&=1+\frac{8\Lambda\alpha_2}{d_1d_2}+\frac{8\alpha_2M}{\Sigma d_2r^{d_1}}-\frac{8\alpha_2q^2}{d_2d_3r^{2d_2}}\\&-\frac{2\alpha_2e^2}{d_2r^{2d_2}}F_1\bigg(\bigg[1,\frac{d_3}{2d_2}\bigg],\bigg[\frac{3d_2-1}{2d_2}\bigg],\frac{-8\zeta q^2(\Gamma(d_1))^2}{r^{2d_2}}\bigg).\label{41a}\end{split}
\end{equation}
The roots of $f_-(r)=0$ offer the horizons of resultant exotic dyonic black holes of Gauss-Bonnet gravity. Hence, using the obtained solution (\ref{37a}), it is easier to verify that the black hole will feature an event horizon when the equation $f_1(r)=0$ is fulfilled for at least one nonzero value of $r$
\begin{eqnarray}\begin{split}
		f_1(r)&=\frac{16\pi G_N M}{d_2\Sigma_{\beta}r^d}-\frac{\beta_1}{r^3}-\frac{\beta_2\alpha_2}{r^5}+\frac{2\Lambda}{d_1d_2r}-\frac{8\pi G_Nq^2}{d_2d_3r^{2d-3}}\\&-\frac{8\pi G_Ne^2}{d_2r^{2d-3}}F_1\bigg(\bigg[1,\frac{d_3}{2d_2}\bigg],\bigg[\frac{3d_2-1}{2d_2}\bigg],\frac{-8\zeta q^2(\Gamma(d_1))^2}{r^{2d_2}}\bigg). \label{41a1}\end{split}
\end{eqnarray}
  Given the constraints $f_-(r_e)=0$ and $f_-^{'}(r_e)=0$, one can figure out the mass of the extreme exotic black hole of Gauss-Bonnet theory. Denoting the event horizon of the extreme black hole by $r_e$, its mass can be derived as
\begin{eqnarray}\begin{split}
		M_e&=\frac{d_2\Sigma_{\beta} r^{d}_e}{32\pi G_Nd_1}\bigg[\frac{\beta_{1(ext)}}{(d+1)^{-1}r_e^3}+\frac{ \beta_{2(ext)}\alpha_2}{(d+3)^{-1}r_e^5}-\frac{2\Lambda}{d_2r_e}+\frac{8\pi G_N e_{ext}^2d_3r_e^{-1}}{d_2(r_e^{2d_2}+8\zeta q_{ext}^2(\Gamma(d_1))^2)}\\&+\frac{16\pi G_Ne_{ext}^2d_1}{d_2r_e^{2d-3}} F_1\bigg(\bigg[1,\frac{d_3}{2d_2}\bigg],\bigg[\frac{3d_2-1}{2d_2}\bigg],\frac{-8\zeta q^2(\Gamma(d_1))^2}{r_e^{2d_2}}\bigg)+\frac{8\pi G_Nq_{ext}^2(3d-5)}{d_2d_3r_e^{2d-3}}\bigg], \label{42a1}\end{split}
\end{eqnarray}
 where $q_{ext}$ and $e_{ext}$ indicate the extreme values of magnetic and electric charges, respectively. In addition, the following equation provides the extreme value of horizon radius
 \begin{eqnarray}\begin{split}
 \frac{d_3\beta_{1(ext)}}{r_e^3}+\frac{d_5\beta_{2(ext)}\alpha_2}{r_e^5}-\frac{2\Lambda}{d_2r_e}-\frac{8\pi G_Nq_{ext}^2}{d_2r_e^{2d-3}}-\frac{8\pi G_Ne_{ext}^2d_3}{d_2r_e(r_e^{2d_2}+8\zeta q_{ext}^2(\Gamma(d_1))^2)}=0. \label{43a1}\end{split}
 \end{eqnarray}
   \begin{figure}[h]
	\centering
	\includegraphics[width=0.8\textwidth]{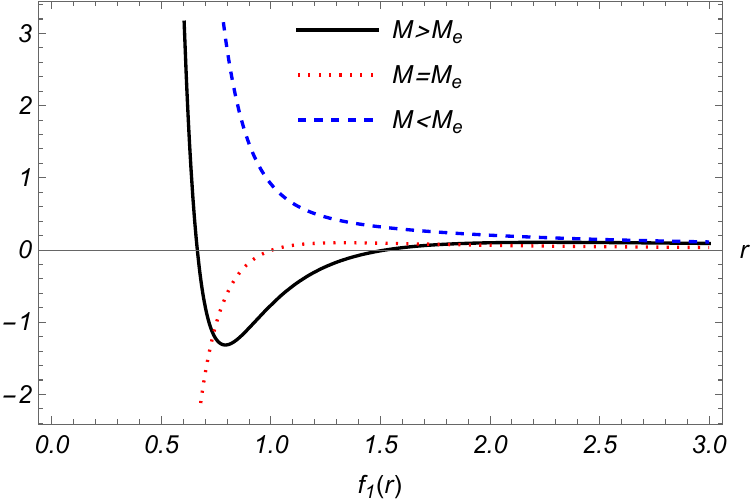}
	\caption{The behaviour of resultant exotic black hole solution (Eq. (\ref{37a})) for different values of mass $M$. The fixed values of the other parameters are taken as $d=5$, $G_N=1$, $\Sigma_{\beta}=1$, $q=0.5$, $\zeta=2$, $e=0.5$ $\beta_1=1$, $\beta_2=1$, $\alpha_2=0.05$ and $\Lambda=-0.1$.}\label{skr4a}
\end{figure} 
\begin{figure}[h]
	\centering
	\includegraphics[width=0.8\textwidth]{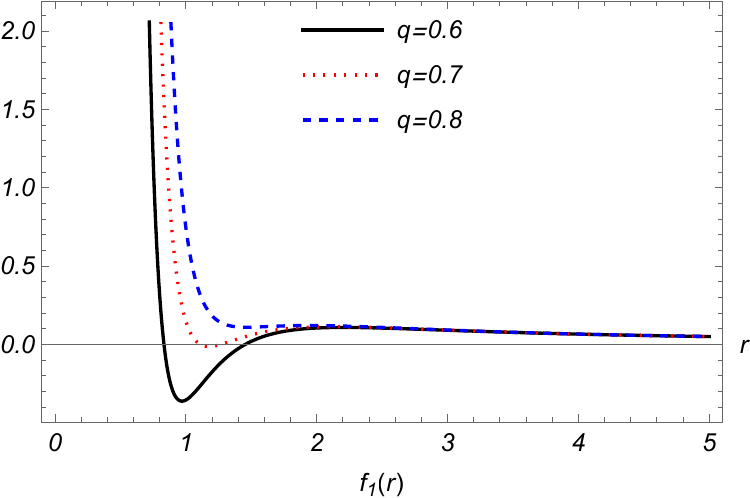}
	\caption{Effect of the magnetic charge $q$ on the horizon structure of exotic black solution (Eq. (\ref{37a})). The fixed values of the other parameters are taken as $d=5$, $G_N=1$, $\Sigma_{\beta}=1$, $M=0.2$, $\zeta=2$, $e=0.5$ $\beta_1=1$, $\beta_2=1$, $\alpha_2=0.05$ and $\Lambda=-0.1$.}\label{skr4b}
\end{figure}  
\begin{figure}[h]
	\centering
	\includegraphics[width=0.8\textwidth]{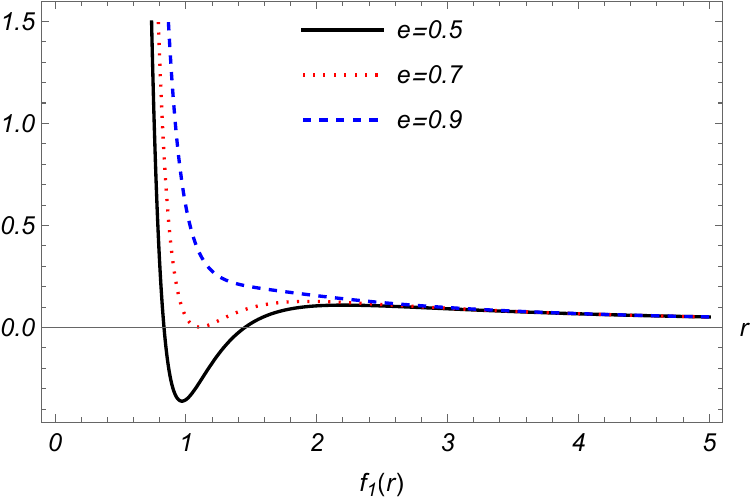}
	\caption{Effect of the electric charge $e$ on the horizon structure of exotic black solution (Eq. (\ref{37a})). The fixed values of the other parameters are taken as $d=5$, $G_N=1$, $\Sigma_{\beta}=1$, $M=0.2$, $\zeta=2$, $q=0.6$ $\beta_1=1$, $\beta_2=1$, $\alpha_2=0.05$ and $\Lambda=-0.1$.}\label{skr4c}
\end{figure} 
\begin{figure}[h]
	\centering
	\includegraphics[width=0.8\textwidth]{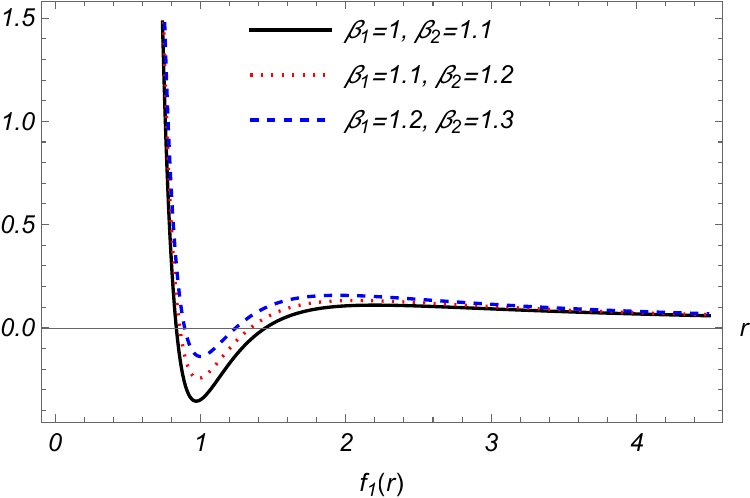}
	\caption{Effect of the topological parameters on the horizon structure of exotic black solution (Eq. (\ref{37a})). The fixed values of the other parameters are taken as $d=5$, $G_N=1$, $\Sigma_{\beta}=1$, $M=0.2$, $\zeta=2$, $q=0.6$, $e=0.5$, $\alpha_2=0.05$ and $\Lambda=-0.1$.}\label{skr4d}
\end{figure} 
\begin{figure}[h]
	\centering
	\includegraphics[width=0.8\textwidth]{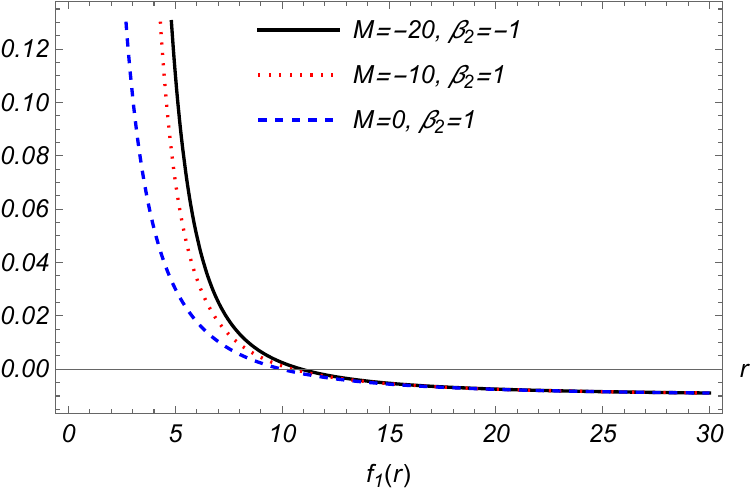}
	\caption{The behaviour of resultant exotic black hole solution (Eq. (\ref{37a})) in de Sitter space. The fixed values of the other parameters are taken as $d=6$, $G_N=1$, $\Sigma_{\beta}=1$, $q=0.5$, $\zeta=2$, $e=0.5$, $\alpha_2=0.05$ and $\Lambda=-0.1$.}\label{skr4e}
\end{figure} 

 Fig. \ref{skr4a} portrays the response of the resulting exotic dyonic solution (\ref{37a}) for multiple values of the black hole's mass $M$. The horizons of exotic black holes with electric charge $e$ and magnetic charge $q$ in Gauss-Bonnet gravity appear at the positions where the respective curve surpasses the horizontal axis. It is seen that the solution will illustrate the object with a single event horizon when the mass takes its extreme value $M=M_{e}$. Similarly, the black hole features an inner (Cauchy) horizon $r_-$ and an outer (event) horizon $r_+$ when $M>M_e$. Moreover, the solution that comes out will reflect the naked singularity whenever mass of black hole satisfies $M<M_e$. In a similar fashion, Figs. \ref{skr4b} and \ref{skr4c} show the responses of our exotic solution according to multiple values of magnetic and electric charges, respectively. It is evident that the radii of inner and outer horizons are getting closer to one another when the magnitudes of $q$ and $e$ are rising. As an outcome, there also exist extreme values $q=q_{ext}$ and $e=e_{ext}$ for which the black hole has a single event horizon. When the magnitudes of the electric and magnetic charges go over these extreme values, the resultant exotic dyonic solution exhibits the naked singularity. Additionally, there will be two horizons for the exotic dyonic black hole whenever $q<q_{ext}$ and $e<e_{ext}$. Similarly, the consequences of topological parameters on our derived solution can be noticed in Fig. \ref{skr4d}. We observe that there also exist a critical values of topological parameters $\beta_i$'s with $i=1,2$ such that the exotic black hole is extreme when $\beta_i=\beta_{i(ext)}$ and the black hole has two horizons when $\beta_i<\beta_{i(ext)}$. In addition, the scenario of naked singularity will manifest itself when $\beta_i>\beta_{i(ext)}$. Note that we have gone with $\Lambda<0$ in the preceding investigation of exotic dyonic black holes. Lastly, an overview of exotic dyonic black hole solutions in de Sitter space is presented in Fig. \ref{skr4e}. One can see that the resultant dyonic solution (\ref{37a}) exhibits an event horizon for any value of $M$ even when the mass parameter vanishes or becomes negative. It is important to note that the Kretschmann scalar (\ref{36a1}) is also singular at $r=0$ when $f(r)$ is replaced by (\ref{37a}). Hence, we have a true curvature singularity at $r=0$.

\subsection{Thermodynamics of exotic dyonic black holes in Gauss-Bonnet gravity}
 Here, we desire to work out some fundamental thermodynamic quantities of the exotic dyonic black holes in Gauss-Bonnet gravity. With the assumption $f_-(r_+)=0$, it is effortless to pinpoint the finite mass of the exotic dyonic black hole as 
\begin{eqnarray}\begin{split}
		M&=\frac{d_2\Sigma_{\beta} r^{d_1}_+}{16\pi G_N}\bigg[\frac{\beta_1}{r_+^2}+\frac{\beta_2\alpha_2}{r_+^4}-\frac{2\Lambda}{d_1d_2}+\frac{8\pi G_Nq^2}{d_2d_3r_+^{2d_2}}\\&+\frac{8\pi G_Ne^2}{d_2r_+^{2d_2}}F_1\bigg(\bigg[1,\frac{d_3}{2d_2}\bigg],\bigg[\frac{3d_2-1}{2d_2}\bigg],\frac{-8\zeta q^2(\Gamma(d_1))^2}{r_+^{2d_2}}\bigg)\bigg]. \label{44a1}\end{split}
\end{eqnarray}
One would possibly compute the expression of Hawking temperature \cite{57} from the theoretical idea of surface gravity as follows:
\begin{equation}
	T_H=\frac{1}{2\pi}\bigg[-\frac{1}{2}(\nabla_{\kappa}\mathcal{X}_{\sigma})(\nabla^{\kappa}\mathcal{X}^{\sigma})\bigg]^{\frac{1}{2}},\label{45a1}
\end{equation}
where $\mathcal{X}_{\sigma}$ symbolizes the temporal Killing vector field. Thus, the Hawking temperature $T_H$ adopts the following form
\begin{eqnarray}\begin{split}
		T_{H}&=\frac{1}{4\pi(r_+^2+2\alpha_2\beta_1)}\bigg[d_3\beta_1r_++\frac{d_5\alpha_2\beta_2}{r_+}-2\Lambda r_+^3-\frac{8\pi G_Nq^2}{d_2r_+^{2d_3-1}}\\&-\frac{8\pi G_Nd_3e^2r_+^3}{d_2\big(r_+^{2d_2}+8\zeta q^2(\Gamma(d_1))^2\big)}\bigg].\label{46a1}\end{split} 
\end{eqnarray}
\begin{figure}[h]
	\centering
	\includegraphics[width=0.8\textwidth]{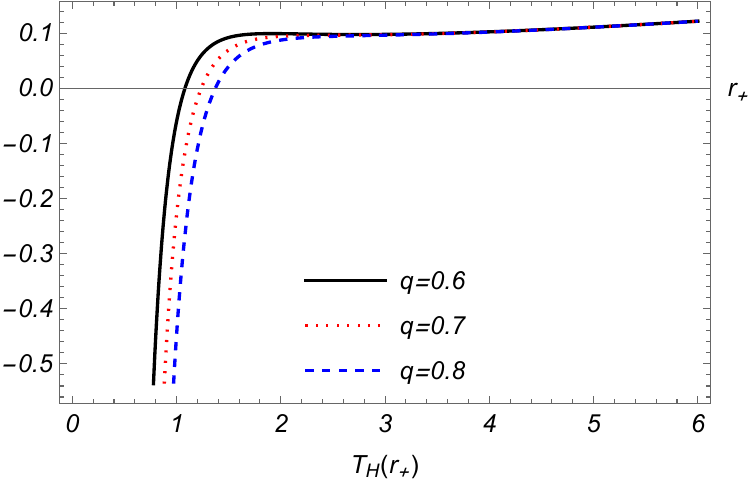}
	\caption{The Hawking temperature $T_H$ (Eq. (\ref{46a1})) is plotted for various values of magnetic charge. The other parameters have the following values: $d=5$, $G_N=1$, $\Sigma_{\beta}=1$, $e=0.5$, $\zeta=2$, $\alpha_2=0.05$, $\beta_1=1$, $\beta_2=1.1$ and $\Lambda=-0.1$.}\label{skr5a}
\end{figure}
\begin{figure}[h]
	\centering
	\includegraphics[width=0.8\textwidth]{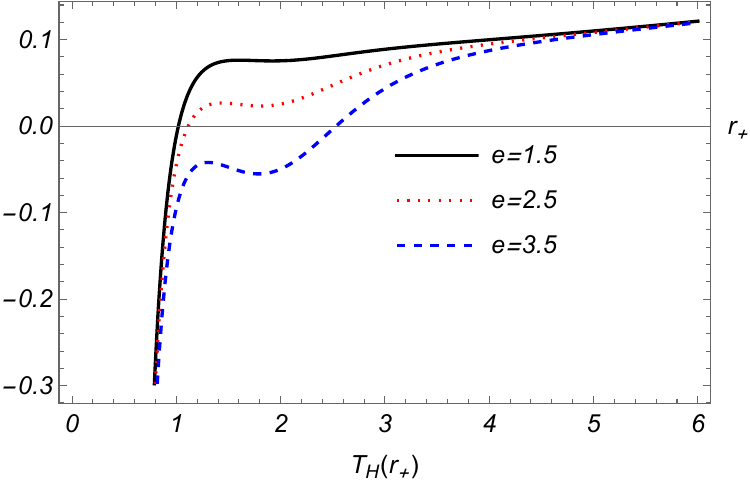}
	\caption{The Hawking temperature $T_H$ (Eq. (\ref{46a1})) is plotted for various values of electric charge. The other parameters have the following values: $d=5$, $G_N=1$, $\Sigma_{\beta}=1$, $q=0.5$, $\zeta=2$, $\alpha_2=0.05$, $\beta_1=1$, $\beta_2=1.1$ and $\Lambda=-0.1$.}\label{skr5b}
\end{figure}
\begin{figure}[h]
	\centering
	\includegraphics[width=0.8\textwidth]{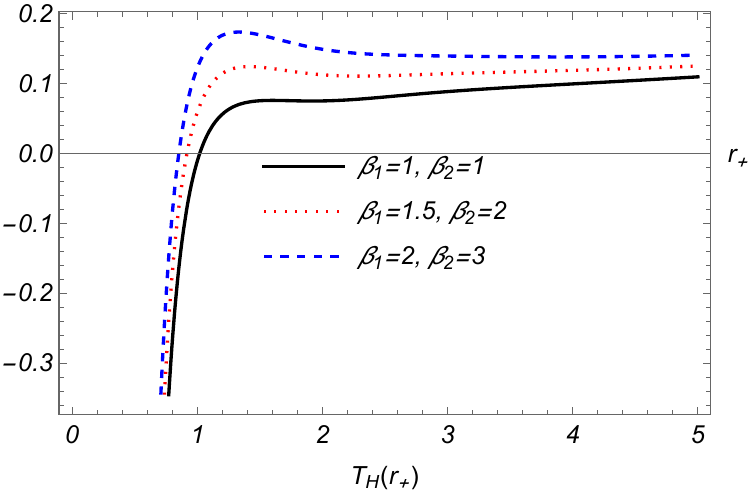}
	\caption{Impact of the topological parameters on the behaviour of the temperature $T_H$ (Eq. (\ref{46a1})). The specific values i.e. $d=5$, $G_N=1$, $\Sigma_{\beta}=1$, $q=0.5$, $e=1.5$, $\zeta=2$, $\alpha_2=0.05$ and $\Lambda=-0.1$ are considered for the other parameters.}\label{skr5c}
\end{figure}

 The variation of Hawking temperature (\ref{46a1}) for multiple amounts of the electromagnetic charges and topological parameters is displayed in Figs. \ref{skr5a}-\ref{skr5c}, respectively. The horizon radius of extreme black hole is apparent by the value at which $T_H$ vanishes. As the magnitudes of electromagnetic charges increase, it is apparent that the size of extreme black hole grows. The size of this object, however, diminishes when the topological parameters grow together. In addition, the black hole whose event horizon radius falls within the domain of positive temperature suggests that it would be physical. 

Since the solution (\ref{37a}) establishes the exotic dyonic black holes of Gauss-Bonnet gravity or second order Lovelock gravity, the area law is incapable of being fully implemented \cite{58,59}. As a result, the determination of this quantity can benefit greatly from taking advantage of Hamiltonian formalism \cite{60,61,62}. This formalism has additionally been employed in the circumstances involving higher curvature theories and Wald's approach is implemented for estimating the entropy \cite{63,64}. Thus, the entropy of the exotic Lovelock black hole can be assessed as
\begin{equation}\begin{split}
		S&=\frac{d_2\Sigma_{\beta}}{4G_N}\sum_{p=1}^{p_{max}}\frac{p\beta_{p-1}\alpha_pr_+^{d_{2p}}}{d_{2p}}. \label{47a1}\end{split}
\end{equation}
It is effortless to work out the entropy of exotic dyonic Gauss-Bonnet black hole from the preceding generic expression of entropy (\ref{47a1}). This will be carried out by setting $p_{max}=2$, which makes the expression for the entropy as 
\begin{equation}\begin{split}
		S&=\frac{d_2\Sigma_{\beta}}{4G_N}\bigg(\frac{r_+^{d_2}}{d_2}+\frac{2\beta_1\alpha_2r_+^{d_4}}{d_4}\bigg). \label{48a1}\end{split}
\end{equation}
The electric and magnetic charges are readily derived from the fluxes of $F_{[2]}$ and $H_{[d_2]}$ at infinity, respectively. They are, in fact, formulated as follows:
\begin{align}\begin{split}
		&e\sim \int\ast F_{[2]} ,\\&
		q\sim\int H_{[d_2]},\label{49a1}\end{split}
\end{align} 
with effective proportionality constants. By considering the re-scaled Lovelock coupling parameters $\alpha_p$'s, entropy $S$, magnetic charge $q$, and electric charge $Q$ as extensive thermodynamic variables, it becomes possible to extend the first law of thermodynamics as
\begin{equation}
	dM=T_HdS+\mathcal{A}de+\mathcal{U}dq-\frac{1}{16\pi G_N}\sum_{p=0}^{2}\tilde{\Psi}^{(p)}d\alpha_p,\label{50a1}
\end{equation} 
where $T_H=\big(\frac{\partial M}{\partial S}\big)\vert_{q,e,\alpha_p}$. In addition, the conjugate potentials affiliated with the Lovelock coupling parameters $a_p$'s are determined from Eqs. (\ref{44a1}) and (\ref{48a1}) in the following way:
\begin{equation}\begin{split}
		\Psi^{(p)}&=\frac{d_2\Sigma_{\beta}r_+^{d_{2p}}}{16\pi G_N}\bigg[\frac{\beta_{p}}{r_+}-\frac{4\pi T_Hp\beta_{p-1}}{d_{2p}}\bigg].\label{51a1}\end{split}
\end{equation}
The thermodynamic volume can be found by placing $p=0$ in Eq. (\ref{51a1}). Consequently, we have
\begin{equation}
	V=-\tilde{\Psi}^{(0)}=\frac{16\pi G_N\Psi^{(0)}}{d_1d_2}=\frac{\Sigma_{\beta}r_+^{d_1}}{d_1},\label{52a1}
\end{equation} 
in which $\tilde{\Psi}^{(0)}$ is a conjugate quantity associated with the re-scaled parameter $\alpha_0$. Note that the above thermodynamic volume is conjugated to the thermodynamical pressure 
\begin{equation}
	P=-\frac{\Lambda}{8\pi}=\frac{d_1d_2\alpha_0}{16\pi}.\label{53a1}
\end{equation}
It is vital to note that this thermodynamical pressure is only effective when $\Lambda<0$ i.e. $\alpha_0>0$. Since we had earlier chosen $\alpha_1=1$, the entity $\tilde{\Psi}^{(1)}$ will disappear. Fortunately, the conjugate quantity $\tilde{\Psi}^{(2)}$ affiliated with the re-scaled Lovelock parameter $\alpha_2$ can be acquired as
\begin{equation}
\tilde{\Psi}^{(2)}=\frac{\Sigma_{\beta}d_2}{16\pi G_Nd_3d_4}r_+^{d_4}\bigg(\frac{\beta_2}{r_+}-\frac{8\pi T_H\beta_1}{d_4}\bigg).\label{54a1}	
\end{equation}
Moreover, the electric potential $\mathcal{A}$ takes the form
\begin{equation}
	\mathcal{A}=\frac{e\Sigma_{\beta}}{r_+^{d_3}}F_1\bigg(\bigg[1,\frac{d_3}{2d_2}\bigg],\bigg[\frac{3d_2-1}{2d_2}\bigg],\frac{-8\zeta q^2(\Gamma(d_1))^2}{r_+^{2d_2}}\bigg),\label{55a1}
\end{equation}
whereas the magnetic potential can be worked out as
\begin{equation}\begin{split}
		\mathcal{U}&=\frac{q\Sigma_{\beta}}{d_3r_+^{d_3}}+\frac{d_3e^2\Sigma_{\beta}r_+^{d_1}}{d_2q\big(r_+^{2d_2}+8\zeta q^2(\Gamma(d_1))^2\big)}\\&-\frac{d_3e^2\Sigma_{\beta}}{d_2qr_+^{d_3}}F_1\bigg(\bigg[1,\frac{d_3}{2d_2}\bigg],\bigg[\frac{3d_2-1}{2d_2}\bigg],\frac{-8\zeta q^2(\Gamma(d_1))^2}{r_+^{2d_2}}\bigg).\label{56a1}\end{split}
\end{equation}
From Eqs. (\ref{52a1})-(\ref{54a1}), one can rewrite the extended first law in a more explicit form as
\begin{equation}
	dM=T_HdS+VdP-\frac{\tilde{\Psi}^{(2)}}{16\pi G_N}d\alpha_2+\mathcal{A}de+\mathcal{U}dq.\label{57a1}
\end{equation}
Associated with this law, Smarr's relation \cite{70a,70b} satisfied by the thermodynamic quantities of exotic Gauss-Bonnet black hole can be constructed as
\begin{equation}
d_3M=d_2T_HS+2PV+\frac{\tilde{\Psi}^{(2)}\alpha_2}{8\pi G_N}+d_3\mathcal{A}e+d_3\mathcal{U}q.\label{58a1}	
\end{equation}
One way to define the specific heat capacity is through  
\begin{equation}
	C_H=T_{H}\frac{dS}{dT_H}\bigg|_{q,Q}. \label{59a1}
\end{equation}
Finally, by using Eq. (\ref{46a1}), the heat capacity can be displayed as 
\begin{equation}\begin{split}
C_H(r_+)&=\frac{\Sigma_{\beta}d_2(r_+^2+2\alpha_2\beta_1)(r_+^{d_3}+2\beta_1\alpha_2r_+^{d_5})\Delta_1(r_+)}{4G_N\bigg[(r_+^2+2\alpha_2\beta_1)\Delta_1'(r_+)-2r_+\Delta_1(r_+)\bigg]},\label{59a}\end{split}
\end{equation}
 where
 \begin{equation}\begin{split}
 \Delta_1(r_+)&=d_3\beta_1r_++\frac{d_5\alpha_2\beta_2}{r_+}-2\Lambda r_+^3-\frac{8\pi G_Nq^2}{d_2r_+^{2d-7}}-\frac{8\pi G_Ne^2d_3r_+^3}{d_2\big(r_+^{2d_2}+8\zeta q^2(\Gamma(d_1))^2\big)},\label{60a}\end{split}
 \end{equation}
 and 
\begin{equation}\begin{split}
\Delta_1'(r_+)&=d_3\beta_1-\frac{d_5\alpha_2\beta_2}{r_+^2}-6\Lambda r_+^2+\frac{8\pi G_N(2d_3-1)q^2}{d_2r_+^{2d_3}}-\frac{24\pi G_Nd_3e^2r_+^2}{d_2\big(r_+^{2d_2}+8\zeta q^2(\Gamma(d_1))^2\big)}\\&+\frac{16\pi G_Nd_3e^2r_+^{2d_1}}{d_2\big(r_+^{2d_2}+8\zeta q^2(\Gamma(d_1))^2\big)^2}.\label{61a}\end{split}
\end{equation}
\begin{figure}[h]
	\centering
	\includegraphics[width=0.8\textwidth]{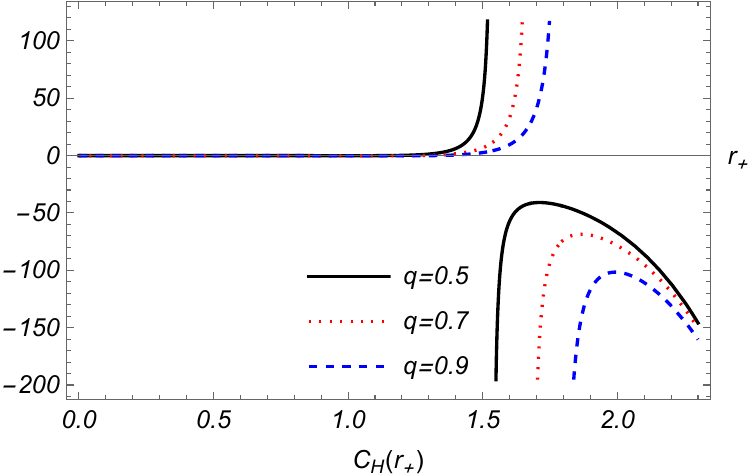}
	\caption{Dependence of heat capacity (Eq. (\ref{59a})) on the event horizon for multiple values of magnetic charge. The other parameters are selected as $d=5$, $G_N=1$, $\Sigma_{\beta}=1$, $e=1$, $\zeta=0.5$, $\alpha_2=0.05$, $\beta_1=0.1$, $\beta_2=0.1$ and $\Lambda=-0.01$.}\label{skr6a}
\end{figure}
\begin{figure}[h]
	\centering
	\includegraphics[width=0.8\textwidth]{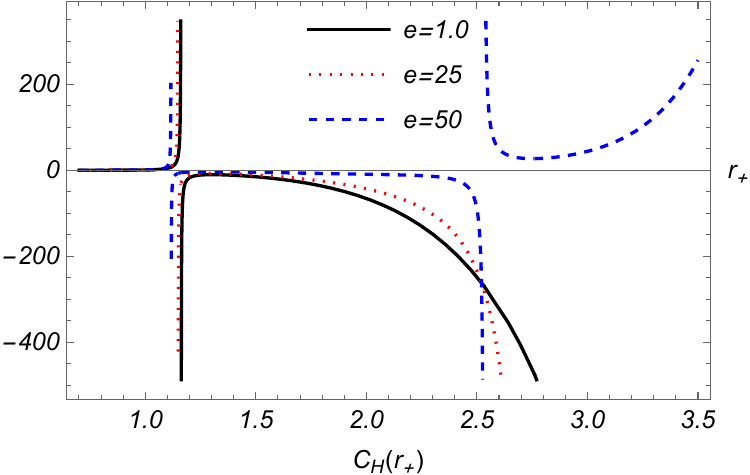}
	\caption{Dependence of heat capacity (Eq. (\ref{59a})) on the event horizon for multiple values of electric charge. The other parameters are selected as $d=7$, $G_N=1$, $\Sigma_{\beta}=1$, $q=0.5$, $\zeta=2$, $\alpha_2=0.05$, $\beta_1=1$, $\beta_2=1$ and $\Lambda=-0.1$.}\label{skr6b}
\end{figure}
\begin{figure}[h]
	\centering
	\includegraphics[width=0.8\textwidth]{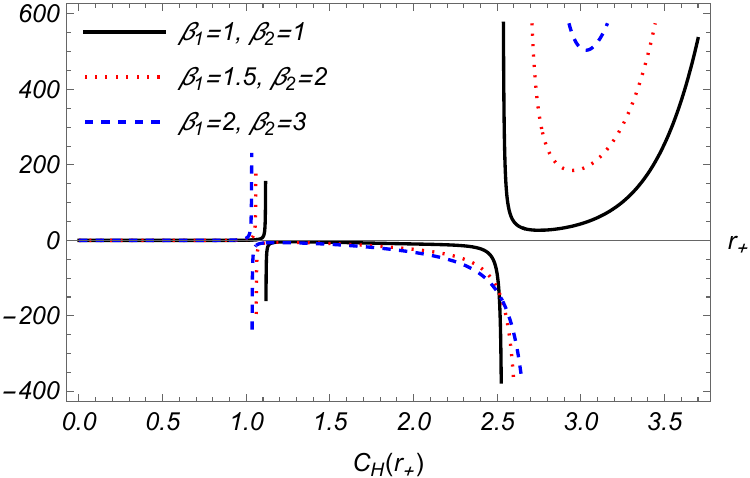}
	\caption{Impact of the topological parameters on the heat capacity (Eq. (\ref{59a})) with $d=7$, $G_N=1$, $\Sigma_{\beta}=1$, $q=0.5$, $\zeta=2$, $\alpha_2=0.05$, $e=50$ and $\Lambda=-0.1$.}\label{skr6c}
\end{figure}

The consequences of the electromagnetic and topological parameters on the local stability of exotic dyonic black holes of Gauss-Bonnet gravity are presented in Figs. \ref{skr6a}-\ref{skr6c}. The local stability of the black hole (\ref{37a}) is linked to a positive heat capacity, whereas the instability is connected to the negativity of this quantity. The locations at which this quantity becomes singular refer to the second-order phase transition, whereas the spots at which it is zero correspond to the first-order phase transition. We have revealed that there exist two values of event horizon radius i.e $r_1$ and $r_2$ at which $C_H$ is infinite. Likewise, there also exist a value $r_c$ at which $C_H$ vanishes. It is noticeable that $C_H$ is negative when event horizon radius falls in the interval $(0,r_c)$. Therefore, the black hole with an event horizon satisfying $0<r_+<r_c$ remains locally unstable. Meanwhile, the black hole would be stable whenever the associated $r_+$ belongs to $(r_c,r_1)$. Furthermore, $C_H$ is yet again negative in $(r_1,r_2)$ so this also reflects the unstable zone for our exotic dyonic black holes. Ultimately, all the exotic dyonic black holes would be locally stable when their event horizons lie in $(r_2,\infty)$. Additionally, when the magnetic charge is growing in magnitude, the points of phase transitions are also increasing. However, when electric charge advances, $r_c$ increases, whereas $r_1$ and $r_2$ diminish. Similarly, when the topological parameters acquire higher values, $r_c$ and $r_1$ drop, while $r_2$ rises.

\section{Exotic dyonic black holes of the third order Lovelock gravity}

In this part, we explore the exciting exotic dyonic black holes of third order Lovelock gravity and their thermodynamic properties. The resulting Lovelock polynomial (\ref{33a1}) offers the following following form for $p_{max}=3$
\begin{eqnarray}\begin{split}
		&-\frac{\alpha_3f^3}{r^6}+\bigg(\frac{\alpha_2}{r^4}+\frac{3\beta_1\alpha_3}{r^6}\bigg)f^2-\bigg(\frac{1}{r^2}+\frac{2\beta_1\alpha_2}{r^4}+\frac{3\beta_2\alpha_3}{r^6}\bigg)f(r)\\&-\frac{2\Lambda}{d_1d_2}+\frac{\beta_1}{r^2}+\frac{\beta_2\alpha_2}{r^4}+\frac{\beta_3\alpha_3}{r^6}=\mathcal{N}(r),\label{62a1}\end{split}
\end{eqnarray}
where $\mathcal{N}(r)$ is stated as
\begin{equation}
	\mathcal{N}(r)=\frac{16\pi G_NM}{d_2\Sigma_{\beta}r^{d_1}}-\frac{8\pi G_Nq^2}{d_2d_3r^{2d_2}}-\frac{8\pi G_Ne^2}{d_2r^{2d_2}} F_1\bigg(\bigg[1,\frac{d_3}{2d_2}\bigg],\bigg[\frac{3d_2-1}{2d_2}\bigg],\frac{-8\zeta q^2(\Gamma(d_1))^2}{r^{2d_2}}\bigg).\label{63a1}
\end{equation}
The real solution to the preceding third degree polynomial equation can be worked out as
 \begin{eqnarray}\begin{split}
 		f(r)&=\frac{1}{3\alpha_3\big(\frac{1}{4}\big(\Xi+\sqrt{\gamma}\big)\big)^{1/3}}\bigg[\bigg(r^4(3\alpha_3+\alpha_2^2)-9(\beta_1^2-\beta^2)\alpha_3^2\bigg)\\&+(r^2\alpha_2+3\alpha_3\beta_1)\bigg(\frac{1}{2}\big(\Xi+\sqrt{\gamma}\big)\bigg)^{1/3}-\bigg(\frac{1}{2}\big(\Xi+\sqrt{\gamma}\big)\bigg)^{2/3}\bigg],\label{64a1}\end{split}
 \end{eqnarray}
 where
 \begin{equation}
 	\Xi=\big(81\beta_1\beta_2-54\beta_1^3-27\beta_3\big)\alpha_3^3+\bigg[27\alpha_3^2\bigg(\mathcal{N}(r)+\frac{2\Lambda}{d_1d_2}\bigg)+9\alpha_2\alpha_3-2\alpha_2^3\bigg]r^6,\label{65a1}
 \end{equation}
 and
 \begin{equation}\begin{split}
 	\gamma&=\bigg[27\alpha_3^3\bigg(2\beta_1^3-3\beta_1\beta_2+\beta_3\bigg)-\bigg(27\alpha_3^2\bigg(\mathcal{N}(r)+\frac{2\Lambda}{d_1d_2}\bigg)-2\alpha_2^3+9\alpha_2\alpha_3\bigg)r^6\bigg]^2\\&+4\bigg(9\alpha_3^2(\beta_2-\beta_1^2)-(\alpha_2^2+3\alpha_3)r^4\bigg)^3.\label{66a1}\end{split}
 \end{equation}
It needs to be mentioned that the solution provided by Eq. (\ref{64a1}) is the sole physically relevant solution. The physically valid solution would be the one that offers an analogous solution of ETG when the parameters $\alpha_2$ and $\alpha_3$ gradually approach zero. However, the other two roots of Eq. (\ref{62a1}) do not reduce to the solutions of ETG as they are complex conjugate roots. In order to fulfill the criteria for offering an asymptotically anti-de Sitter solution, a constraint must be set for the selection of Lovelock coupling parameters $\alpha_2$ and $\alpha_3$ \cite{9f3,71b}. This restriction can be identified from the positivity of the discriminant of Eq. (\ref{62a1}) in the vicinity of $r=\infty$. Hence,
\begin{equation}
	\frac{108\Lambda^2\alpha_3^2}{d_1^2d_2^2}-\frac{8\Lambda\alpha_2^3}{d_1d_2}+\frac{36\Lambda\alpha_2\alpha_3}{d_1d_2}-\alpha_2^2+4\alpha_3\leq0.\label{67a1}
\end{equation}
 This inequality imposes an upper (and occasionally lower) bound on the thermodynamical pressure for appropriate choices of the coupling parameters. Thus, we are left with
 \begin{equation}
 	P_{\pm}=\frac{9d_1d_2\alpha_2\alpha_3-2\alpha_2^3\pm2\big(\alpha_2^2-3\alpha_3\big)^{3/2}}{432\pi \alpha_3^2}.\label{68a1}
 \end{equation}
By choosing $\alpha=\alpha_2/\sqrt{\alpha_3}$ and $\tilde{p}=4\sqrt{\alpha_3}P$, Eq. (\ref{68a1}) can be written as
\begin{equation}
	\tilde{p}_{\pm}=\frac{d_1d_2}{108\pi}\bigg(9\alpha-2\alpha^3\pm (\alpha^2-3)^{3/2}\bigg).\label{69a1}
\end{equation}
 One can confirm that both branches terminate when $\alpha=\sqrt{3}$. Observe that the pressure turns imaginary when $\alpha<\sqrt{3}$. This behaviour will eliminate the nature of asymptotically anti-de Sitter spacetime. Furthermore, Eq. (\ref{69a1}) demonstrates that the pressure will achieve its maximum when $\alpha=2$. Hence, there would certainly be no lower bound on the pressure when $\alpha\geq2$. Likewise, the pressure incorporates the lower and upper bounds when $\alpha$ corresponds to $\sqrt{3}<\alpha<2$. 
 
 Additionally, if one selects $p_{max}=3$, $8\pi G_N=1$ and $\alpha_3=\alpha_2^2/3$ in Eq. (\ref{35a1}), the solution of the third order Lovelock gravity that reflects the dyonic black holes with constant curvature horizons \cite{9f9} can be constructed as
\begin{eqnarray}\begin{split}
f(r)&=\xi+\frac{r^2}{\alpha_2}\bigg[1- \big(\mathcal{W}_2(r)\big)^{\frac{1}{3}}\bigg], \label{70a1}\end{split}
\end{eqnarray}
where
 \begin{equation}\begin{split}
 \mathcal{W}_2(r)&=1+\frac{6\alpha_2\Lambda}{d_1d_2}+\frac{3\alpha_2M}{\Sigma d_2r^{d_1}}-\frac{3\alpha_2q^2}{2d_2d_3r^{2d_2}}\\&-\frac{3\alpha_2Q^2}{2d_2r^{2d_2}}F_1\bigg(\bigg[1,\frac{d_3}{2d_2}\bigg],\bigg[\frac{3d_2-1}{2d_2}\bigg],\frac{-8\zeta q^2(\Gamma(d_1))^2}{r^{2d_2}}\bigg).\label{71a1}\end{split}
 \end{equation}
 
 The horizons of this exotic dyonic black hole are easily characterized from the zeros of the resultant metric function. Hence, using Eq. (\ref{64a1}), it is simple to demonstrate that the black hole will have an event horizon if the equation $f_1(r)=0$ is satisfied such that
 \begin{eqnarray}\begin{split}
 		f_1(r)&=\frac{16\pi G_N M}{d_2\Sigma_{\beta}r^d}-\frac{\beta_1}{r^3}-\frac{\beta_2\alpha_2}{r^5}-\frac{\beta_3\alpha_3}{r^6}+\frac{2\Lambda}{d_1d_2r}-\frac{8\pi G_Nq^2}{d_2d_3r^{2d-3}}\\&-\frac{8\pi G_Ne^2}{d_2r^{2d-3}}F_1\bigg(\bigg[1,\frac{d_3}{2d_2}\bigg],\bigg[\frac{3d_2-1}{2d_2}\bigg],\frac{-8\zeta q^2(\Gamma(d_1))^2}{r^{2d_2}}\bigg). \label{71af1}\end{split}
 \end{eqnarray}
 The constraints $f(r_e)=0$ and $f^{'}(r_e)=0$ can also be used to compute the mass of the extreme exotic black hole. Thus, we are able to derive
 \begin{eqnarray}\begin{split}
 		M_e&=\frac{d_2\Sigma_{\beta} r^{d}_e}{32\pi G_Nd_1}\bigg[\frac{\beta_{1(ext)}(d+1)}{r_e^3}+\frac{(d+3)\beta_{2(ext)}\alpha_2}{r_e^5}+\frac{(d+5)\beta_{3(ext)}\alpha_3}{r_e^7}-\frac{2\Lambda}{d_2r_e}
 		\\&+\frac{8\pi G_Nq_{ext}^2(3d-5)}{d_2d_3r_e^{2d-3}}+\frac{16\pi G_Ne_{ext}^2d_1}{d_2r_e^{2d-3}} F_1\bigg(\bigg[1,\frac{d_3}{2d_2}\bigg],\bigg[\frac{3d_2-1}{2d_2}\bigg],\frac{-8\zeta q^2(\Gamma(d_1))^2}{r_e^{2d_2}}\bigg)
 		\\&+\frac{8\pi G_N e_{ext}^2d_3r_e^{-1}}{d_2(r_e^{2d_2}+8\zeta q_{ext}^2(\Gamma(d_1))^2)}\bigg]. \label{71f1}\end{split}
 \end{eqnarray}
 Correspondingly, the extreme value of horizon radius will satisfy the equation
 \begin{eqnarray}\begin{split}
 		\frac{d_3\beta_{1(ext)}}{r_e^3}+\frac{d_5\alpha_2\beta_{2(ext)}}{r_e^5}+\frac{d_7\alpha_3\beta_{3(ext)}}{r_e^7}-\frac{2\Lambda}{d_2r_e}\\&=\frac{8\pi G_Nq_{ext}^2}{d_2r_e^{2d-3}}+\frac{8\pi G_Ne_{ext}^2d_3}{d_2r_e(r_e^{2d_2}+8\zeta q_{ext}^2(\Gamma(d_1))^2)}. \label{71f2}\end{split}
 \end{eqnarray}
 \begin{figure}[h]
 	\centering
 	\includegraphics[width=0.8\textwidth]{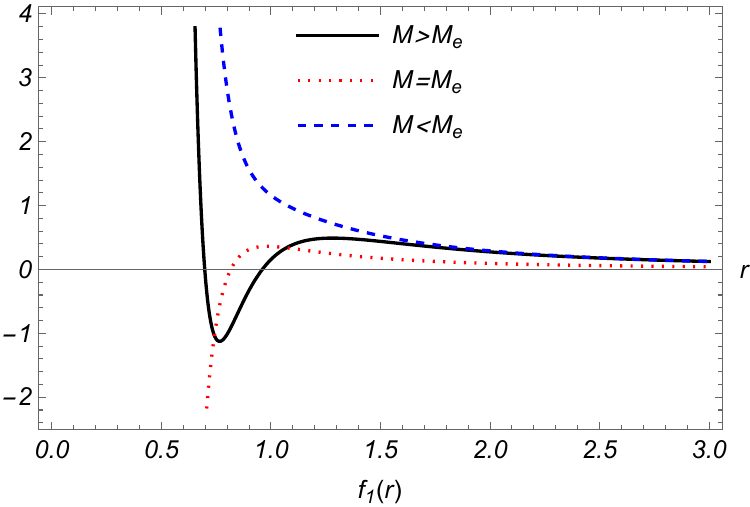}
 	\caption{The behaviour of resultant exotic black hole solution (Eq. (\ref{64a1})) of third order Lovelock gravity for different values of mass $M$. The fixed values of the other parameters are considered as $d=7$, $G_N=1$, $\Sigma_{\beta}=1$, $q=0.5$, $\zeta=2$, $e=0.5$, $\beta_1=1$, $\beta_2=1$, $\beta_3=1$, $\alpha_2=2\sqrt{3\alpha_3}$, $\alpha_3=0.05$ and $\Lambda=-0.1$.}\label{skr7a}
 \end{figure} 
 \begin{figure}[h]
 	\centering
 	\includegraphics[width=0.8\textwidth]{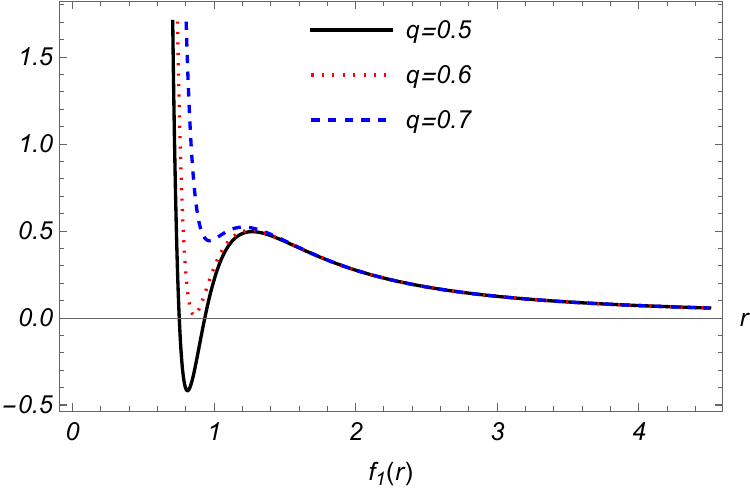}
 	\caption{Effect of the magnetic charge $q$ on the solution (Eq. (\ref{64a1})). Rest of the parameters are fixed as $d=7$, $G_N=1$, $\Sigma_{\beta}=1$, $M=0.2$, $\zeta=2$, $e=0.5$, $\beta_1=1$, $\beta_2=1$, $\beta_3=1$, $\alpha_2=2\sqrt{3\alpha_3}$, $\alpha_3=0.05$ and $\Lambda=-0.1$.}\label{skr7b}
 \end{figure}  
 \begin{figure}[h]
 	\centering
 	\includegraphics[width=0.8\textwidth]{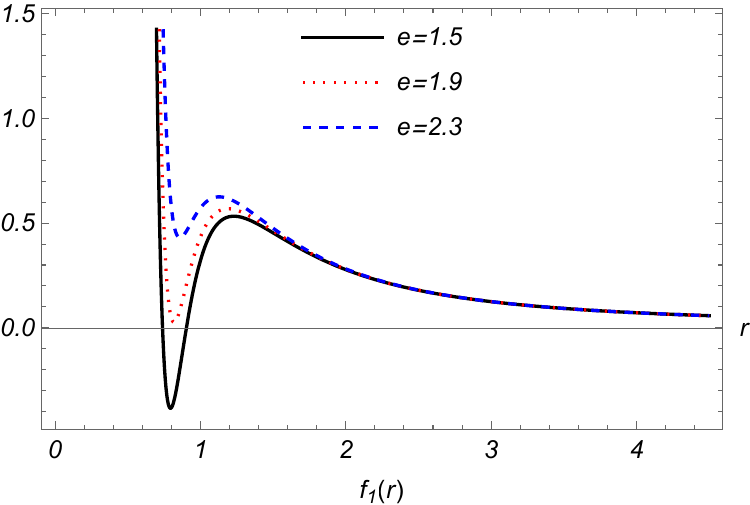}
 	\caption{Effect of the electric charge $e$ on the solution (Eq. (\ref{64a1})). The other parameters are selected as $d=7$, $G_N=1$, $\Sigma_{\beta}=1$, $M=0.2$, $\zeta=2$, $q=0.5$, $\beta_1=1$, $\beta_2=1$, $\beta_3=1$, $\alpha_2=2\sqrt{3\alpha_3}$, $\alpha_3=0.05$ and $\Lambda=-0.1$.}\label{skr7c}
 \end{figure} 
 \begin{figure}[h]
 	\centering
 	\includegraphics[width=0.8\textwidth]{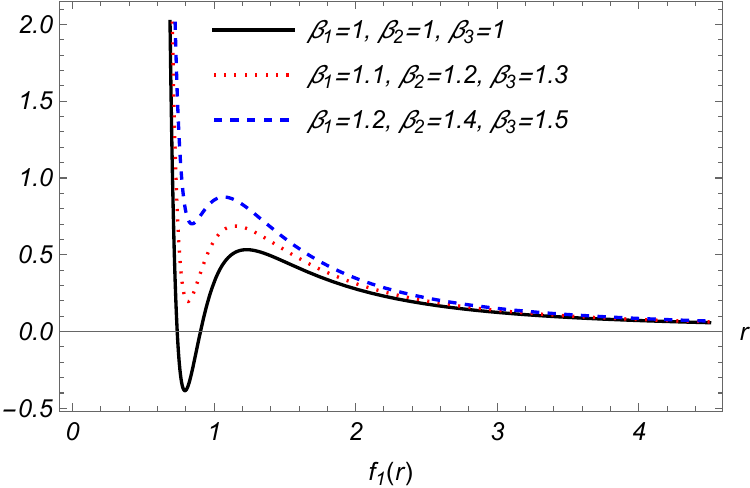}
 	\caption{Effect of the topological parameters on the solution (Eq. (\ref{64a1})). The other parameters are selected as $d=7$, $G_N=1$, $\Sigma_{\beta}=1$, $M=0.2$, $\zeta=2$, $e=1.5$, $q=0.5$, $\alpha_2=2\sqrt{3\alpha_3}$, $\alpha_3=0.05$ and $\Lambda=-0.1$.}\label{skr7d}
 \end{figure}
 \begin{figure}[h]
 	\centering
 	\includegraphics[width=0.8\textwidth]{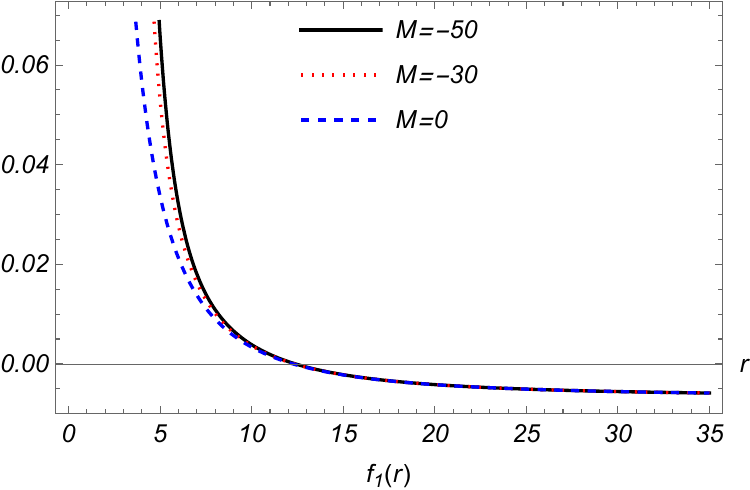}
 	\caption{The behaviour of resultant exotic black hole solution (Eq. (\ref{64a1})) in de Sitter space. The fixed values of the other parameters are considered as $d=7$, $G_N=1$, $\Sigma_{\beta}=1$, $q=0.5$, $\zeta=2$, $e=0.5$, $\alpha_2=\frac{\sqrt{3\alpha_3}}{2}$, $\alpha_3=0.05$ and $\Lambda=0.1$.}\label{skr7e}
 \end{figure}
 
 The effects of mass, electromagnetic charges and topological parameters on the exotic dyonic black hole solution (\ref{64a1}) of the third order Lovelock gravity in anti-de Sitter space are discussed in Figs. \ref{skr7a}-\ref{skr7d}. It is observed that the resultant metric function (\ref{64a1}) can have a single horizon for $M=M_e$, $q=q_{ext}$, $e=e_{ext}$, and $\beta_i's=\beta_{i(ext)}'s$, while it will possess two horizons for $M>M_e$, $q<q_{ext}$, $e<e_{ext}$, and $\beta_i's<\beta_{i(ext)}'s$. Similarly, when $M<M_e$, $q>q_{ext}$, $e>e_{ext}$, and $\beta_i's>\beta_{i(ext)}'s$, the solution will present a naked singularity. Furthermore, when the values of $q$, $e$ and the topological parameters $\beta_i$'s with $i=1,2,3$ are increasing, the inner horizon radius $r_-$ is advancing and the outer horizon radius $r_+$ is dropping. Ultimately, the response of exotic dyonic black hole solutions in de Sitter space is presented in Fig. \ref{skr7e}. As can be noticed, the resultant dyonic exotic solution (\ref{64a1}) with $\Lambda>0$ and $\alpha_2<\sqrt{3\alpha_3}$ encounters an event horizon whatever value $M$ takes, even when the mass parameter vanishes or takes negative values. Note that the Kretschmann scalar (\ref{36a1}) gives singularity at $r=0$ when $f(r)$ is replaced by Eq. (\ref{64a1}). Hence, we have an essential singularity at this position.

\subsection{Thermodynamics of exotic dyonic black holes in third order Lovelock gravity}

 Once again, the black hole's mass can be figured out by employing the condition $f(r_+)=0$ in Eq. (\ref{62a1}) as
  \begin{equation}\begin{split}
 M&=\frac{d_2\Sigma_{\beta}r_+^{d_1}}{16\pi G_N}\bigg[\frac{\beta_1}{r_+^2}+\frac{\beta_2\alpha_2}{r_+^4}+\frac{\beta_3\alpha_3}{r_+^6}-\frac{2\Lambda}{d_1d_2}+\frac{8\pi G_Nq^2}{d_2d_3r_+^{2d_2}}\\&+\frac{8\pi G_Ne^2}{d_2r_+^{2d_2}}F_1\bigg(\bigg[1,\frac{d_3}{2d_2}\bigg],\bigg[\frac{3d_2-1}{2d_2}\bigg],\frac{-8\zeta q^2(\Gamma(d_1))^2}{r_+^{2d_2}}\bigg)\bigg].\label{72a1}\end{split}
 \end{equation}
 Additionally, the Hawking temperature of exotic dyonic black hole can be worked out from Eq. (\ref{45a1}) as
 \begin{equation}\begin{split}
 T_H(r_+)&=\frac{1}{4\pi \big(r_+^4+2\beta_1\alpha_2r_+^2+3\beta_3\alpha_3\big)}\bigg[\beta_1d_3r_+^3+\beta_2\alpha_2d_5r_++\frac{\beta_3\alpha_3d_7}{r_+}\\&-\frac{2\Lambda r_+^5}{d_2}-\frac{8\pi G_Nq^2}{d_2r_+^{2d_4-1}}-\frac{8\pi G_Ne^2d_3r_+^5}{d_2\big(r_+^{2d_2}+8\zeta q^2(\Gamma(d_1))^2\big)}\bigg].\label{73a1}\end{split}
 \end{equation}
 \begin{figure}[h]
 	\centering
 	\includegraphics[width=0.8\textwidth]{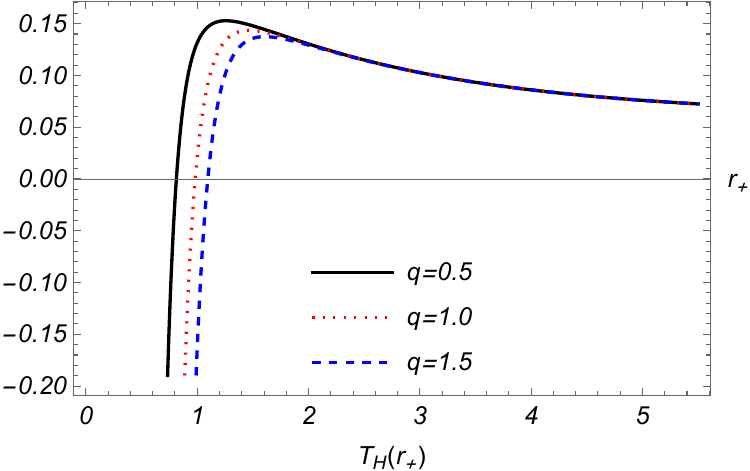}
 	\caption{The Hawking temperature $T_H$ (Eq. (\ref{73a1})) is plotted for various values of magnetic charge. The other parameters have the following values: $d=7$, $G_N=1$, $\Sigma_{\beta}=1$, $e=0.5$, $\zeta=2$, $\alpha_2=2\sqrt{3\alpha_3}$, $\alpha_3=0.05$, $\beta_1=1$, $\beta_2=1.1$, $\beta_3=1.5$, and $\Lambda=-0.1$.}\label{skr8a}
 \end{figure}
 \begin{figure}[h]
 	\centering
 	\includegraphics[width=0.8\textwidth]{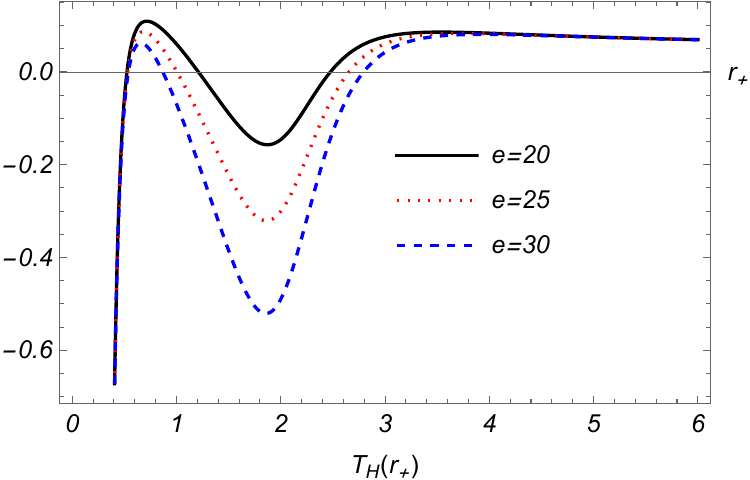}
 	\caption{The Hawking temperature $T_H$ (Eq. (\ref{73a1})) is plotted for various values of electric charge. The other parameters have the following values: $d=7$, $G_N=1$, $\Sigma_{\beta}=1$, $q=0.1$, $\zeta=2$, $\alpha_2=2\sqrt{3\alpha_3}$, $\alpha_3=0.05$, $\beta_1=1$, $\beta_2=1$, $\beta_3=1$, and $\Lambda=-0.1$.}\label{skr8b}
 \end{figure}
 \begin{figure}[h]
 	\centering
 	\includegraphics[width=0.8\textwidth]{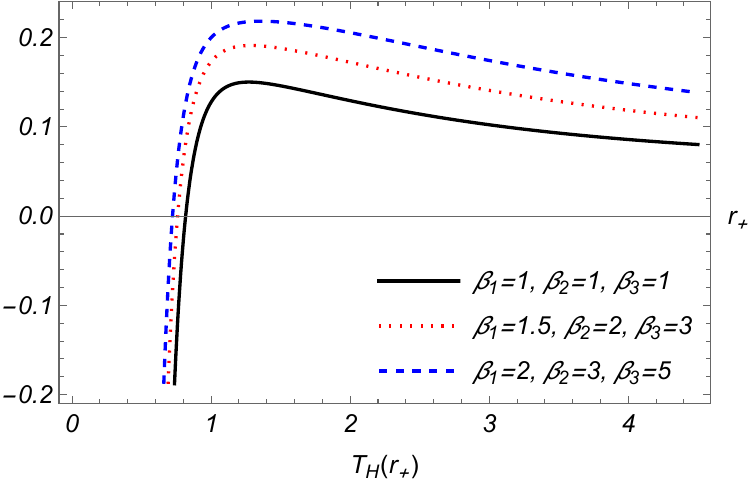}
 	\caption{Impact of the topological parameters on the behaviour of the temperature $T_H$ (Eq. (\ref{73a1})). The specific values i.e. $d=7$, $G_N=1$, $\Sigma_{\beta}=1$, $e=0.5$, $\zeta=2$, $\alpha_2=2\sqrt{3\alpha_3}$, $\alpha_3=0.05$, $q=0.5$ and $\Lambda=-0.1$ are considered for the other parameters.}\label{skr8c}
 \end{figure}
Figs. \ref{skr8a}-\ref{skr8c} display the variation of Hawking temperature (\ref{73a1}) as a function of event horizon. The extreme black hole's size is seen to grow as the intensity of magnetic charge increases. However, when $\beta_i$'s are simultaneously getting bigger, the size of extreme black hole diminishes. It is also noted that the temperature of exotic black holes with higher magnitude of electric charge have three different zeroes, say $r_a$, $r_b$ and $r_c$. Since the physicality of a black hole corresponds to the positivity of $T_H$, therefore, the black hole whose event horizon $r_+$ belongs to either $(r_a,r_b)$, or to $(r_c,\infty)$ would be considered physical. However, the black hole would not be physical when $r_+$ belongs to either $(0,r_a)$, or to $(r_b,r_c)$. Besides, as the parameter $e$ rises, the values of $r_a$ and $r_c$ increase whereas $r_b$ decreases. Additionally, we note that the electromagnetic charges only alter the Hawking temperature of smaller black holes, whereas their influence on the temperature of larger black holes is quite negligible.

 By incorporating $p_{max}=3$, $\beta_0=1$ and $\alpha_1=1$ in Eq. (\ref{47a1}), one will get the entropy of the exotic dyonic black hole solution (\ref{64a1}) as follows:
 \begin{equation}\begin{split}
 S&=\frac{\Sigma_{\beta}d_2}{4G_N}\bigg(\frac{r_+^{d_2}}{d_2}+\frac{2\beta_1\alpha_2r_+^{d_4}}{d_4}+\frac{3\beta_2\alpha_3r_+^{d_6}}{d_6}\bigg).\label{74a1}\end{split}
 \end{equation}
Utilizing the same steps of Section 3, the extended first law associated with Eq. (\ref{64a1}) takes the form
\begin{equation}
	dM=T_HdS+VdP-\frac{\tilde{\Psi}^{(2)}}{16\pi G_N}d\alpha_2-\frac{\tilde{\Psi}^{(3)}}{16\pi G_N}d\alpha_3+\mathcal{A}de+\mathcal{U}dq,\label{75a1}
\end{equation}
where the quantities $V$, $P$, $\tilde{\Psi}^{(2)}$, $\mathcal{A}$ and $\mathcal{U}$ are given by Eqs. (\ref{52a1})-(\ref{56a1}), respectively. Additionally, the conjugate potential $\tilde{\Psi}^{(3)}$ associated with $\alpha_3$ is calculated as
 \begin{equation}\begin{split}
\tilde{\Psi}^{(3)}=\frac{\Sigma_{\beta}d_2}{16\pi G_Nd_3d_4d_5d_6}r_+^{d_6}\bigg(\frac{\beta_3}{r_+}-\frac{12\pi T_H\beta_2}{d_6}\bigg)	.\label{76a1}\end{split}
\end{equation}	
Similarly, Smarr's relation for exotic dyonic black hole will take the form
\begin{equation}
	d_3M=d_2T_HS+2PV+\frac{\tilde{\Psi}^{(2)}\alpha_2}{8\pi G_N}+\frac{\tilde{\Psi}^{(3)}\alpha_3}{4\pi G_N}+d_3\mathcal{A}e+d_3\mathcal{U}q.\label{77a1}	
\end{equation}
 Finally, the heat capacity associated with Eq. (\ref{64a1}) will be derived as
 \begin{equation}\begin{split}
 C_H(r_+)&=\frac{d_2\Sigma_{\beta}\mathcal{E}_1(r_+)(r_+^4+2\beta_1\alpha_2r_+^2+3\beta\alpha_3)(r_+^{d_3}+2\beta_1\alpha_2r_+^{d_5}+3\beta_2\alpha_3r_+^{d_7})}{4G_N\big((r_+^4+2\beta_1\alpha_2r_+^2+3\beta\alpha_3)\mathcal{E}_1'(r_+)-(4r_+^3+4\beta_1\alpha_2r_+)\mathcal{E}_1(r_+)\big)},\label{78a1}\end{split}
 \end{equation}
 with
 \begin{equation}\begin{split}
 \mathcal{E}_1(r_+)&=\beta_1d_3r_+^3+\beta_2\alpha_2d_5r_++\frac{\beta_3\alpha_3d_7}{r_+}-\frac{2\Lambda r_+^5}{d_2}\\&-\frac{8\pi G_Nq^2}{d_2r_+^{2d_4-1}}-\frac{8\pi G_Ne^2d_3r_+^5}{d_2(r_+^{2d_2}+8\zeta q^2(\Gamma(d_1))^2)},\label{79a1}\end{split}
 \end{equation}
 and 
 \begin{equation}\begin{split}
 \mathcal{E}_1'(r_+)&=3\beta_1d_3r_+^2+\beta_2\alpha_2d_5-\frac{\beta_3\alpha_3d_7}{r_+^2}-\frac{10\Lambda r_+^4}{d_2}+\frac{8\pi G_Nq^2(2d-9)}{d_2r_+^{2d_4}}\\&-\frac{40\pi G_Ne^2d_3r_+^4}{d_2(r_+^{2d_2}+8\zeta q^2(\Gamma(d_1))^2)}+\frac{16\pi G_Ne^2d_3r_+^{2d}}{(r_+^{2d_2}+8\zeta q^2(\Gamma(d_1))^2)^2}.\label{80a1}\end{split}
 \end{equation}
 \begin{figure}[h]
 	\centering
 	\includegraphics[width=0.8\textwidth]{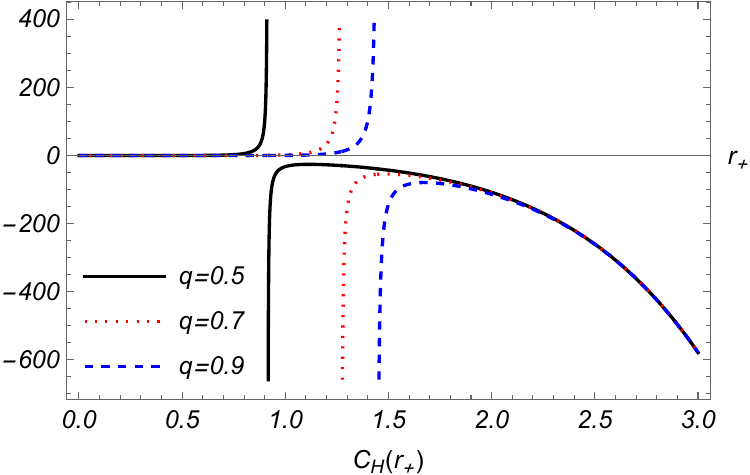}
 	\caption{The heat capacity $C_H$ (Eq. (\ref{78a1})) is plotted for various values of magnetic charge. The other parameters have the following values: $d=7$, $G_N=1$, $\Sigma_{\beta}=1$, $e=0.5$, $\zeta=2$, $\alpha_2=2\sqrt{3\alpha_3}$, $\alpha_3=0.05$, $\beta_1=1$, $\beta_2=1$, $\beta_3=1$, and $\Lambda=-0.1$.}\label{skr9a}
 \end{figure}
 \begin{figure}[h]
 	\centering
 	\includegraphics[width=0.8\textwidth]{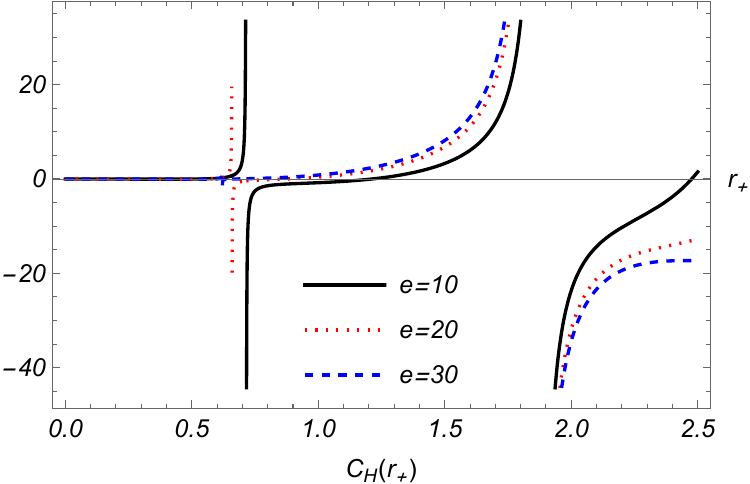}
 	\caption{The heat capacity $C_H$ (Eq. (\ref{78a1})) is plotted for various values of magnetic charge. The other parameters have the following values: $d=7$, $G_N=1$, $\Sigma_{\beta}=1$, $q=0.1$, $\zeta=2$, $\alpha_2=2\sqrt{3\alpha_3}$, $\alpha_3=0.05$, $\beta_1=1$, $\beta_2=1.1$, $\beta_3=1.5$, and $\Lambda=-0.1$.}\label{skr9b}
 \end{figure}
 \begin{figure}[h]
 	\centering
 	\includegraphics[width=0.8\textwidth]{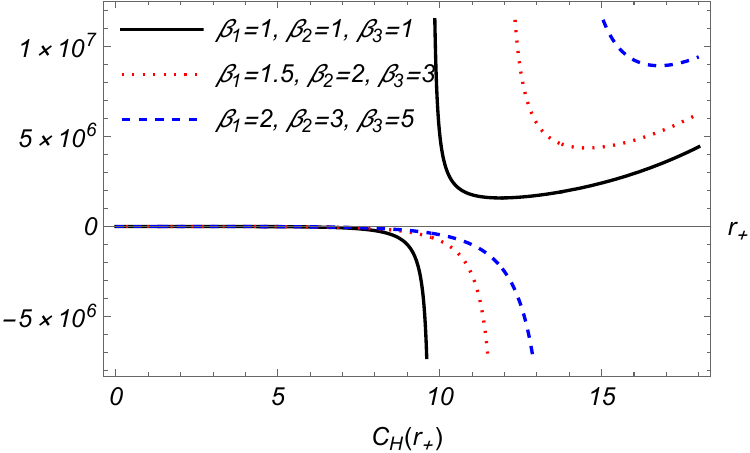}
 	\caption{Impact of the topological parameters on the behaviour of the heat capacity $C_H$ (Eq. (\ref{78a1})). The specific values i.e. $d=7$, $G_N=1$, $\Sigma_{\beta}=1$, $e=0.5$, $\zeta=2$, $\alpha_2=2\sqrt{3\alpha_3}$, $\alpha_3=0.05$, $q=0.5$ and $\Lambda=-0.1$ are selected for the other parameters.}\label{skr9c}
 \end{figure}
 
The consequences of the electromagnetic and topological parameters on the local stability of exotic dyonic black holes of third order Lovelock gravity are presented in Figs. \ref{skr9a}-\ref{skr9c}. We have noticed from the behaviour of $C_H$ that there exist three values of $r_+$ i.e $r_1$, $r_2$ and $r_3$ at which $C_H$ is infinite. Additionally, there also exist values $r_c$, $r_d$ and $r_e$ at which $C_H$ vanishes. Note that $r_1<r_2<r_3$ and $r_c<r_d<r_e$. It is concluded that $C_H$ is negative when $r_+$ falls in any of the intervals $(0,r_c)$, $(r_1,r_d)$, and $(r_2,r_e)$. Therefore, the black hole with event horizon belonging to any of these intervals is locally unstable. On the other side, the black hole would be stable when the associated $r_+$ belongs to either $(r_c,r_1)$ or to $(r_d,r_2)$ or to $(r_e,r_3)\cup(r_3,\infty)$. Additionally, when the electric charge is growing in magnitude, the points $r_c$ and $r_e$ increase and $r_b$ decreases. However, when the magnetic charge is on the rise, only $r_c$ increases, whereas $r_d$ and $r_e$ decrease. Further, the electromagnetic charges have also produced unavoidable influence on the positions of second-order points. For instance, $r_1$ and $r_2$ decrease and $r_3$ increases when both $q$ and $e$ grow. Moreover, when the topological parameters attain higher values, all the transition points decrease except $r_3$ which increases. Finally, we have also observe that larger black holes i.e. objects with event horizon $r_+>r_3$ are more locally stable than the smaller black holes.

\section{Summary and Conclusion}

Lovelock gravity as a logical extension of ETG that authorizes higher dimensional black hole solutions. This theory also allows for black holes which are distinguished by nonconstant curvature horizon geometries. In other words, they feature exotic horizon cross-sections. This stands in a clear opposition to the vacuum solutions of four dimensional Einstein's theory where Birkhoff's theorem puts a restriction of transverse metrics isometric to $\mathcal{S}^2$ for the case of spherically symmetric geometry. In the present work, we addressed the exotic dyonic black holes of Lovelock gravity and their thermodynamic properties. Note that we have used the extended model of QT electromagnetism as a matter source of Lovelock gravity. First we introduced the field equations of Lovelock gravity sourced by extended QT electromagnetism where a transverse part of $d$-dimensional spacetime is a nonconstant curvature base manifold. Then we found the Lovelock polynomial which satisfies the equations of motion and can also generate the exotic dyonic black hole solutions of the Lovelock gravity with arbitrary order.

 We specifically used the resultant polynomial to work out the exotic dyonic black hole solutions of Gauss-Bonnet gravity. To serve this purpose, we figure out the metric function (\ref{37a}) which relies on the mass, electromagnetic charges, and topological parameters $\beta_i$'s with $i=1,2$. It is concluded that this solution can represent an exotic dyonic black hole that have a single horizon for $M=M_e$, $q=q_{ext}$, $e=e_{ext}$, and $\beta_i's=\beta_{i(ext)}'s$. Similarly, it represents the exotic dyonic black hole with two horizons when the inequalities $M>M_e$, $q<q_{ext}$, $e<e_{ext}$, and $\beta_i's<\beta_{i(ext)}'s$ are satisfied. Besides, the solution will represent naked singularity when $M<M_e$, $q>q_{ext}$, $e>e_{ext}$, and $\beta_i's>\beta_{i(ext)}'s$. Additionally, we have also computed the thermodynamic quantities as functions of the event horizon. The generalized first law and Smarr's formula associated with the resultant black holes of Gauss-Bonnet gravity are also constructed. It is significant to recognize that the thermodynamic mass, Hawking temperature and heat capacity are influenced by the electromagnetic charges and topological parameters. However, there is no direct influence of the electromagnetic charges on the entropy. Thus, as regards the entropy of Lovelock black holes, the extended QT electromagnetic field operates identically to Maxwellian and Born-Infeld type electromagnetic fields. Since the Cauchy and event horizons depend on these charges through Eq. (\ref{44a1}), they are indirectly affecting the entropy. To investigate local stability, we plotted the temperature (\ref{46a1}) and heat capacity (\ref{59a}) for various values of $q$, $e$ and topological parameters. Regarding this, Figs. \ref{skr5a}-\ref{skr5c} demonstrated that the extreme black hole grows in size when the values of electromagnetic charges increase. However, it decreases when the magnitudes of topological parameters increase. The exotic dyonic black hole of Gauss-Bonnet gravity specified by solution (\ref{37a}) with event horizon $r_+$ will be locally stable when the heat capacity is positive. In contrast, those values that preserve an agreement with the negative $C_H$ indicate the outer horizons of unstable black holes. The plots of $C_H$ additionally brought us to the conclusion that there exist zeros and singularities of this quantity in its domain. Consequently, both first and second order phase transitions are taking place for the resultant objects. In light of this, Figs. \ref{skr6a}-\ref{skr6c} revealed that there exist two values of the event horizon radius i.e $r_1$ and $r_2$ at which $C_H$ is infinite. Likewise, there also exist a value $r_c$ at which $C_H$ vanishes. Hence, our conclusion is that the black hole remains locally unstable as long as its event horizon is falling in the interval $(0,r_c)$. On the other side, the black hole with the associated $r_+$ in $(r_c,r_1)$ is locally stable. Furthermore, $C_H<0$ in $(r_1,r_2)$, so this interval also corresponds to the region of instability for our exotic dyonic black hole. Additionally, we have observed that the exotic dyonic black holes are locally stable when the associated event horizons fall in $(r_2,\infty)$. It is also concluded that the changes in the magnitudes of electromagnetic charges and topological parameters strongly impact the sizes of the intervals associated with local stability and instability.
 
 Moreover, we have also investigated the exotic dyonic black holes of the third order Lovelock theory. In this setup, we computed the metric function (\ref{64a1}) which describes the exotic dyonic black holes in higher dimensions. Then we studied the effects of mass, electromagnetic charges and topological parameters on these exotic dyonic black holes in anti-de Sitter space. Figs. \ref{skr7a}-\ref{skr7d} showed that the resultant metric function (\ref{64a1}) represents a black hole with a single horizon for $M=M_e$, $q=q_{ext}$, $e=e_{ext}$, and $\beta_i's=\beta_{i(ext)}'s$ and describes black hole with two horizons for $M>M_e$, $q<q_{ext}$, $e<e_{ext}$, and $\beta_i's<\beta_{i(ext)}'s$. Similarly, naked singularity appeared for the situation when $M<M_e$, $q>q_{ext}$, $e>e_{ext}$, and $\beta_i's>\beta_{i(ext)}'s$. In addition, we analyzed that the Cauchy horizon $r_-$ grows whereas the event horizon $r_+$ falls when the values of $q$, $e$ and $\beta_i$'s with $i=1,2,3$ increase. Finally, we presented the behaviour of exotic dyonic black hole solutions in de Sitter space (Fig. \ref{skr7e}). It is examined that the resultant exotic dyonic solution (\ref{64a1}) of third order Lovelock gravity with $\Lambda>0$ and $\alpha_2<\sqrt{3\alpha_3}$ possesses an event horizon for any value of $M$ even when the mass parameter vanishes or takes negative values. To examine thermodynamic behaviour of these black holes in anti-de Sitter space, we have also derived the expressions of thermodynamic quantities including, temperature, entropy and heat capacity. We also derived the generalized forms of the first law and Smarr's formula for these black holes of third order Lovelock gravity. Figs. \ref{skr8a}-\ref{skr8c} showed that the extreme black hole's size increases with the increase in the magnitude of magnetic charge. However, the size of this black hole decreases when the topological parameters increase simultaneously. We have also noticed that the temperature of exotic black holes with higher magnitude of electric charge vanishes at three different points, i.e. $r_a$, $r_b$ and $r_c$. Hence, we concluded that the exotic dyonic black hole whose event horizon $r_+$ belongs to either $(r_a,r_b)$, or to $(r_c,\infty)$ would be physically valid. The black hole would, however, not be physical when $r_+$ belongs to either $(0,r_a)$ or  $(r_b,r_c)$. It is also observed that the electromagnetic charges have a significant effect on the Hawking temperature of smaller black holes, whereas their influence on $T_H$ of larger black holes is very small. The regions of local stability of the exotic dyonic black holes of third order Lovelock gravity are investigated in Figs. \ref{skr9a}-\ref{skr9c}. We observed that there exist three values of $r_+$ i.e. $r_1$, $r_2$ and $r_3$ at which $C_H$ is infinite. Additionally, this quantity vanishes at three different points $r_c$, $r_d$ and $r_e$. We concluded that the exotic dyonic black hole given by Eq. (\ref{64a1}) is locally unstable when $r_+$ falls in any of the intervals $(0,r_c)$, $(r_1,r_d)$, and $(r_2,r_e)$. On the other hand, this black hole is locally stable when the associated $r_+$ belongs to either $(r_c,r_1)$ or $(r_d,r_2)$ or $(r_e,r_3)\cup(r_3,\infty)$.  
 
Exploring the effects of QT electromagnetism on the critical behaviour and greybody factors of our derived black hole solutions (\ref{37a}) and (\ref{64a1}) could be crucial. Moreover, other physical properties of exotic black holes in quasi-topological gravities might also be quite appealing. These concepts have been left for our subsequent work.

\section*{Acknowledgements}
The author KS gratefully acknowledges a research grant from the Abdus Salam International Centre for Theoretical Physics (ICTP), Trieste, Italy, that made possible his visit to Queen Mary University of London, London, during the summer of 2024.

\end{document}